%% file: bare_jrnl_compsoc.tex
\documentclass[10pt,journal,compsoc]{IEEEtran}
\ifCLASSOPTIONcompsoc
  \usepackage[nocompress]{cite}
\else
  \usepackage{cite}
\fi
\ifCLASSINFOpdf
\else
\fi
\usepackage{amsmath}

\usepackage{enumitem}
\usepackage{hyperref}
\usepackage{multirow}
\usepackage{graphicx}
\usepackage{csquotes}
\usepackage{xspace}
\usepackage{xcolor}
\usepackage{float}
\usepackage{svg}
\usepackage{booktabs}
\usepackage{soul}
\usepackage{lipsum}
\usepackage{booktabs}
\usepackage{siunitx}
\usepackage{comment}
\usepackage{tabularx}
\usepackage{placeins}
\usepackage{subfig}
\usepackage{colortbl}
\usepackage{listings}

\input{macros/commands.tex}
\input{macros/constants.tex}
\input{macros/orcid.tex}

\begin{document}
%
% paper title
% Titles are generally capitalized except for words such as a, an, and, as,
% at, but, by, for, in, nor, of, on, or, the, to and up, which are usually
% not capitalized unless they are the first or last word of the title.
% Linebreaks \\ can be used within to get better formatting as desired.
% Do not put math or special symbols in the title.
\title{Can Clean New Code reduce\\Technical Debt Density?}
%
%
% author names and IEEE memberships
% note positions of commas and nonbreaking spaces ( ~ ) LaTeX will not break
% a structure at a ~ so this keeps an author's name from being broken across
% two lines.
% use \thanks{} to gain access to the first footnote area
% a separate \thanks must be used for each paragraph as LaTeX2e's \thanks
% was not built to handle multiple paragraphs
%
%
%\IEEEcompsocitemizethanks is a special \thanks that produces the bulleted
% lists the Computer Society journals use for "first footnote" author
% affiliations. Use \IEEEcompsocthanksitem which works much like \item
% for each affiliation group. When not in compsoc mode,
% \IEEEcompsocitemizethanks becomes like \thanks and
% \IEEEcompsocthanksitem becomes a line break with idention. This
% facilitates dual compilation, although admittedly the differences in the
% desired content of \author between the different types of papers makes a
% one-size-fits-all approach a daunting prospect. For instance, compsoc 
% journal papers have the author affiliations above the "Manuscript
% received ..."  text while in non-compsoc journals this is reversed. Sigh.

\author{George~Digkas \orcidicon{0000-0003-0590-5477},
        Alexander~Chatzigeorgiou \orcidicon{0000-0002-5381-8418},
        Apostolos~Ampatzoglou \orcidicon{0000-0002-5764-7302},
        and~Paris~Avgeriou\orcidicon{0000-0002-7101-0754},~\IEEEmembership{Senior~Member,~IEEE}% <-this % stops a space
\IEEEcompsocitemizethanks{\IEEEcompsocthanksitem G. Digkas is with the Institute of Mathematics and Computer Science,
University of Groningen, Netherlands and with the Department of Applied Informatics, University of Macedonia, Greece% <-this % stops an unwanted space
.\protect\\
% note need leading \protect in front of \\ to get a newline within \thanks as
% \\ is fragile and will error, could use \hfil\break instead.
E-mail: g.digkas@rug.nl, g.digkas@uom.edu.gr
\IEEEcompsocthanksitem A. Chatzigeorgiou is with the Department of Applied Informatics, University of Macedonia, Greece% <-this % stops an unwanted space
.\protect\\
E-mail: achat@uom.edu.gr

\IEEEcompsocthanksitem A. Ampatzoglou is with the Department of Applied Informatics, University of Macedonia, Greece% <-this % stops an unwanted space
.\protect\\
E-mail: a.ampatzoglou@uom.edu.gr

\IEEEcompsocthanksitem P. Avgeriou is with the Institute of Mathematics and Computer Science,
University of Groningen, Netherlands% <-this % stops an unwanted space
.\protect\\
E-mail: p.avgeriou@rug.nl}% <-this % stops an unwanted space
\thanks{Manuscript received April 19, 2005; revised August 26, 2015.}}

% note the % following the last \IEEEmembership and also \thanks - 
% these prevent an unwanted space from occurring between the last author name
% and the end of the author line. i.e., if you had this:
% 
% \author{....lastname \thanks{...} \thanks{...} }
%                     ^------------^------------^----Do not want these spaces!
%
% a space would be appended to the last name and could cause every name on that
% line to be shifted left slightly. This is one of those "LaTeX things". For
% instance, "\textbf{A} \textbf{B}" will typeset as "A B" not "AB". To get
% "AB" then you have to do: "\textbf{A}\textbf{B}"
% \thanks is no different in this regard, so shield the last } of each \thanks
% that ends a line with a % and do not let a space in before the next \thanks.
% Spaces after \IEEEmembership other than the last one are OK (and needed) as
% you are supposed to have spaces between the names. For what it is worth,
% this is a minor point as most people would not even notice if the said evil
% space somehow managed to creep in.

% The paper headers
\markboth{Journal of \LaTeX\ Class Files,~Vol.~14, No.~8, August~2015}%
{Shell \MakeLowercase{\textit{et al.}}: Bare Demo of IEEEtran.cls for Computer Society Journals}
% The only time the second header will appear is for the odd numbered pages
% after the title page when using the twoside option.
% 
% *** Note that you probably will NOT want to include the author's ***
% *** name in the headers of peer review papers.                   ***
% You can use \ifCLASSOPTIONpeerreview for conditional compilation here if
% you desire.

% The publisher's ID mark at the bottom of the page is less important with
% Computer Society journal papers as those publications place the marks
% outside of the main text columns and, therefore, unlike regular IEEE
% journals, the available text space is not reduced by their presence.
% If you want to put a publisher's ID mark on the page you can do it like
% this:
%\IEEEpubid{0000--0000/00\$00.00~\copyright~2015 IEEE}
% or like this to get the Computer Society new two part style.
%\IEEEpubid{\makebox[\columnwidth]{\hfill 0000--0000/00/\$00.00~\copyright~2015 IEEE}%
%\hspace{\columnsep}\makebox[\columnwidth]{Published by the IEEE Computer Society\hfill}}
% Remember, if you use this you must call \IEEEpubidadjcol in the second
% column for its text to clear the IEEEpubid mark (Computer Society jorunal
% papers don't need this extra clearance.)

% use for special paper notices
%\IEEEspecialpapernotice{(Invited Paper)}

% for Computer Society papers, we must declare the abstract and index terms
% PRIOR to the title within the \IEEEtitleabstractindextext IEEEtran
% command as these need to go into the title area created by \maketitle.
% As a general rule, do not put math, special symbols or citations
% in the abstract or keywords.
\IEEEtitleabstractindextext{%
\input{Abstract.tex}

% Note that keywords are not normally used for peerreview papers.
}

% make the title area
\maketitle

% To allow for easy dual compilation without having to reenter the
% abstract/keywords data, the \IEEEtitleabstractindextext text will
% not be used in maketitle, but will appear (i.e., to be "transported")
% here as \IEEEdisplaynontitleabstractindextext when the compsoc 
% or transmag modes are not selected <OR> if conference mode is selected 
% - because all conference papers position the abstract like regular
% papers do.
\IEEEdisplaynontitleabstractindextext
% \IEEEdisplaynontitleabstractindextext has no effect when using
% compsoc or transmag under a non-conference mode.

% For peer review papers, you can put extra information on the cover
% page as needed:
% \ifCLASSOPTIONpeerreview
% \begin{center} \bfseries EDICS Category: 3-BBND \end{center}
% \fi
%
% For peerreview papers, this IEEEtran command inserts a page break and
% creates the second title. It will be ignored for other modes.
\IEEEpeerreviewmaketitle

\input{Introduction.tex}

\input{RelatedWork.tex}

\input{CaseStudyDesign.tex}

\input{Results.tex}

\input{Discussion.tex}

\input{ThreatsToValidity.tex}

\input{Conclusion.tex}

\ifCLASSOPTIONcompsoc
  % The Computer Society usually uses the plural form
  \section*{Acknowledgments}
\else
  % regular IEEE prefers the singular form
  \section*{Acknowledgment}
\fi

Work reported in this paper has received funding from the European Union’s Horizon 2020 research and innovation programme under grant agreement No 780572  (project SDK4ED) and under grant agreement No 801015 (project EXA2PRO).

% Can use something like this to put references on a page
% by themselves when using endfloat and the captionsoff option.
\ifCLASSOPTIONcaptionsoff
  \newpage
\fi

% trigger a \newpage just before the given reference
% number - used to balance the columns on the last page
% adjust value as needed - may need to be readjusted if
% the document is modified later
%\IEEEtriggeratref{8}
% The "triggered" command can be changed if desired:
%\IEEEtriggercmd{\enlargethispage{-5in}}

% references section

% can use a bibliography generated by BibTeX as a .bbl file
% BibTeX documentation can be easily obtained at:
% http://mirror.ctan.org/biblio/bibtex/contrib/doc/
% The IEEEtran BibTeX style support page is at:
% http://www.michaelshell.org/tex/ieeetran/bibtex/
%\bibliographystyle{IEEEtran}
% argument is your BibTeX string definitions and bibliography database(s)
%\bibliography{IEEEabrv,../bib/paper}
%
% <OR> manually copy in the resultant .bbl file
% set second argument of \begin to the number of references
% (used to reserve space for the reference number labels box)
\bibliographystyle{IEEEtran}
\bibliography{references.bib}

% biography section
% 
% If you have an EPS/PDF photo (graphicx package needed) extra braces are
% needed around the contents of the optional argument to biography to prevent
% the LaTeX parser from getting confused when it sees the complicated
% \includegraphics command within an optional argument. (You could create
% your own custom macro containing the \includegraphics command to make things
% simpler here.)
%\begin{IEEEbiography}[{\includegraphics[width=1in,height=1.25in,clip,keepaspectratio]{mshell}}]{Michael Shell}
% or if you just want to reserve a space for a photo:

\begin{IEEEbiography}[{\includegraphics[width=1in,height=1.25in,clip,keepaspectratio]{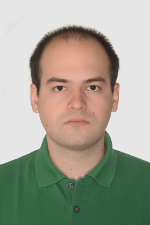}}]{George Digkas}
is a double degree PhD student at the University of Groningen, the Netherlands and the University of Macedonia, Greece. 
He received a BSc and MSc in Applied Informatics from the University of Macedonia, Greece in 2014 and 2016, respectively. His research interests include technical debt, software quality and mining of software repositories. 
\end{IEEEbiography}

\begin{IEEEbiography}[{\includegraphics[width=1in,height=1.25in,clip,keepaspectratio]{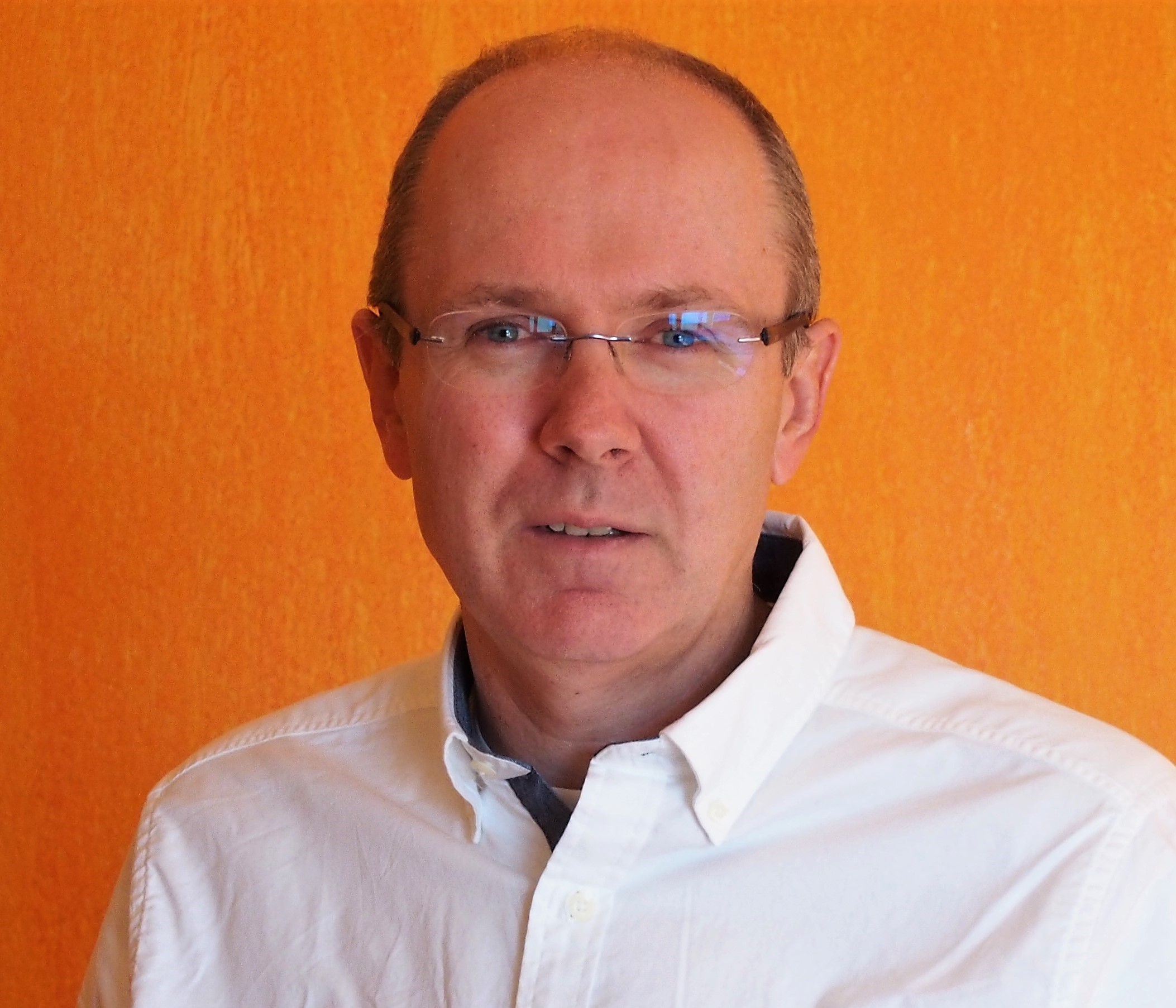}}]{Dr. Alexander Chatzigeorgiou}
is a Professor of Software Engineering in the Department of Applied Informatics and Dean of the School of Information Sciences at the University of Macedonia, Thessaloniki, Greece. He received the Diploma in Electrical Engineering and the PhD degree in Computer Science from the Aristotle University of Thessaloniki, Greece, in 1996 and 2000, respectively. From 1997 to 1999 he was with Intracom S.A., Greece, as a software designer. His research interests include object-oriented design, software maintenance, technical debt and evolution analysis. He has published more than 150 articles in international journals and conferences and participated in a number of European and national research programs. He is a Senior Associate Editor of the Journal of Systems and Software.
\end{IEEEbiography}

\begin{IEEEbiography}[{\includegraphics[width=1in,height=1.25in,clip,keepaspectratio]{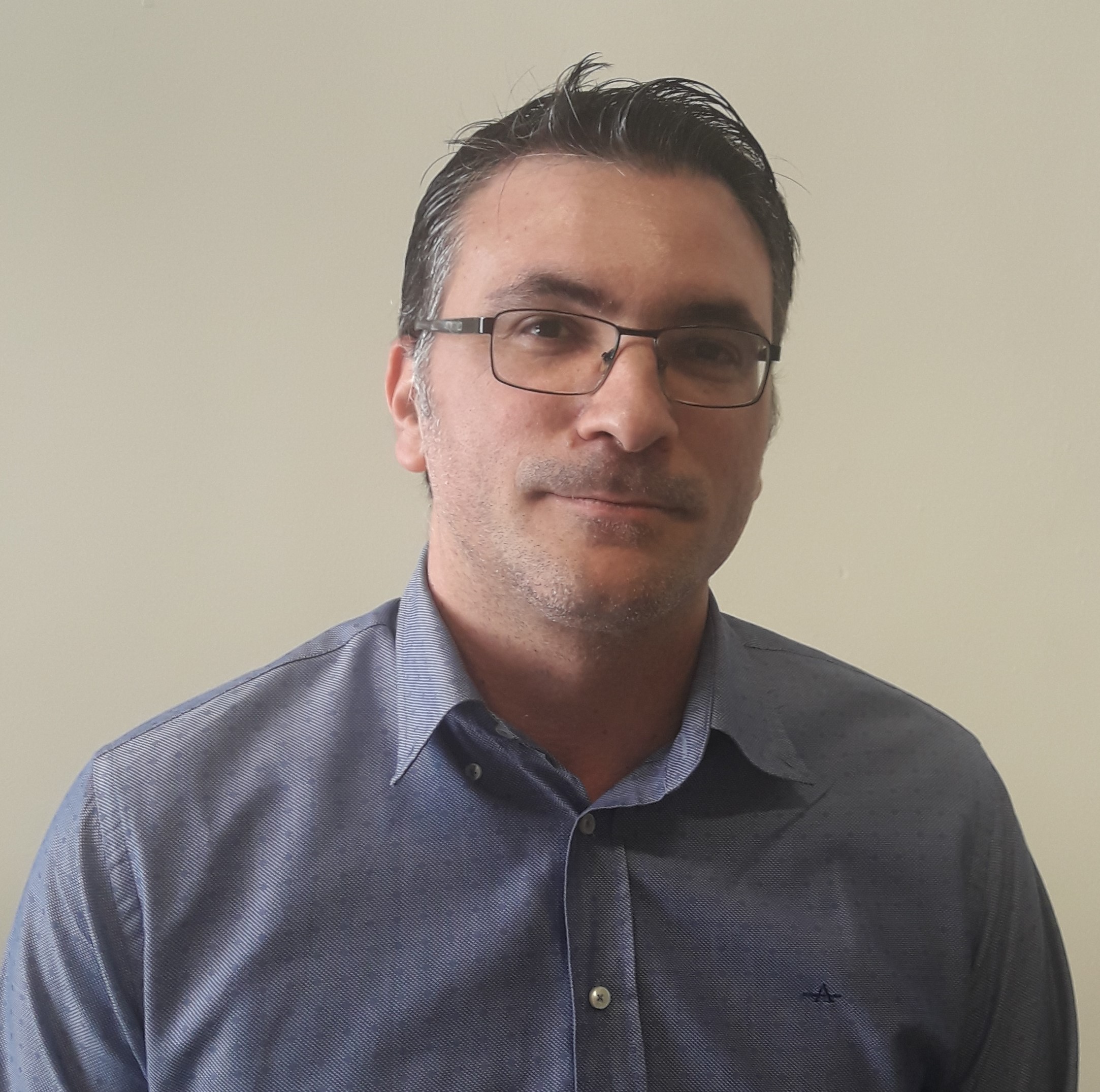}}]{Dr. Apostolos Ampatzoglou}
is an Assistant Professor in the Department of Applied Informatics in University of Macedonia (Greece), where he carries out research and teaching in the area of software engineering. Before joining University of Macedonia he was an Assistant Professor in the University of Groningen (Netherlands). He holds a BSc on Information Systems (2003), an MSc on Computer Systems (2005) and a PhD in Software Engineering by the Aristotle University of Thessaloniki (2012). He has published more than 70 articles in international journals and conferences, and is/was involved in over 15 R\&D ICT projects, with funding from national and international organizations. His current research interests are focused on technical debt management, software maintainability, reverse engineering software quality management, open source software, and software design.
\end{IEEEbiography}

\begin{IEEEbiography}[{\includegraphics[width=1in,height=1.25in,clip,keepaspectratio]{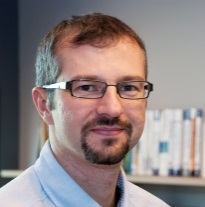}}]{Dr. Paris Avgeriou}
is Professor of Software Engineering in the Johann Bernoulli Institute for Mathematics and Computer Science, University of Groningen, the Netherlands where he has led the Software Engineering research group since September 2006. Before joining Groningen, he was a post-doctoral Fellow of the ERCIM. He has participated in a number of national and European research projects related to the European industry of Software-intensive systems. He has co-organized several international conferences and workshops (mainly at the International Conference on Software Engineering - ICSE). He sits on the editorial board of Springer Transactions on Pattern Languages of Programming (TPLOP). He has edited special issues in IEEE Software, Journal of Systems and Software and Springer TPLOP. He has published more than 130 peer-reviewed articles in international journals, conference proceedings and books. His research interests lie in the area of software architecture, with strong emphasis on architecture modeling, knowledge, evolution, patterns and link to requirements.
\end{IEEEbiography}

% You can push biographies down or up by placing
% a \vfill before or after them. The appropriate
% use of \vfill depends on what kind of text is
% on the last page and whether or not the columns
% are being equalized.

%\vfill

% Can be used to pull up biographies so that the bottom of the last one
% is flush with the other column.
%\enlargethispage{-5in}

% that's all folks
\end{document}

%% file: macros/commands.tex
\newcommand{\project}[1]{\texttt{#1}}
\newcommand\rnd[2]{\sisetup{input-decimal-markers={.}, round-mode=places, round-precision=#1}\hphantom{-\,}\num{#2}}

\newcommand{\squote}[1]{`#1'}
\newcommand{\dquote}[1]{``#1''}

%% file: macros/constants.tex
\newcommand{\sq}{SonarQube\xspace}
\newcommand{\os}{open-source\xspace}
\newcommand{\oss}{open-source software\xspace}
\newcommand{\TD}{Technical Debt\xspace}
\newcommand{\td}{technical debt\xspace}
\newcommand{\TDM}{Technical Debt Management\xspace}
\newcommand{\pom}{pom.xml\xspace}
\newcommand{\asf}{Apache Software Foundation\xspace}
\newcommand{\TDd}{TD$_{density}$\xspace}
\newcommand{\TDI}{Technical Debt Issue\xspace}
\newcommand{\TDIs}{Technical Debt Issues\xspace}

\newcommand{\rqone}{\textit{Among the three major types of code changes (insertion, deletion and modification) which is primarily responsible for changes in Technical Debt density?}\xspace}
\newcommand{\rqtwo}{\textit{Is the frequency of commits in which new code is cleaner compared to exiting code, associated with the existence of practices targeting code quality?}\xspace}
\newcommand{\ECSAsyscount}{sixty-six\xspace}
\newcommand{\TSEsyscount}{twenty-seven\xspace}
\newcommand{\TSEsyscountNumber}{27\xspace}
\newcommand{\TSEclasses}{500\xspace}
\newcommand{\TSEcommits}{1000\xspace}
\newcommand{\TSEyears}{3\xspace}

%% file: macros/orcid.tex
\usepackage{scalerel}
\usepackage{tikz}
\usetikzlibrary{svg.path}

\definecolor{orcidlogocol}{HTML}{A6CE39}
\tikzset{
  orcidlogo/.pic={
    \fill[orcidlogocol] svg{M256,128c0,70.7-57.3,128-128,128C57.3,256,0,198.7,0,128C0,57.3,57.3,0,128,0C198.7,0,256,57.3,256,128z};
    \fill[white] svg{M86.3,186.2H70.9V79.1h15.4v48.4V186.2z}
                 svg{M108.9,79.1h41.6c39.6,0,57,28.3,57,53.6c0,27.5-21.5,53.6-56.8,53.6h-41.8V79.1z M124.3,172.4h24.5c34.9,0,42.9-26.5,42.9-39.7c0-21.5-13.7-39.7-43.7-39.7h-23.7V172.4z}
                 svg{M88.7,56.8c0,5.5-4.5,10.1-10.1,10.1c-5.6,0-10.1-4.6-10.1-10.1c0-5.6,4.5-10.1,10.1-10.1C84.2,46.7,88.7,51.3,88.7,56.8z};
  }
}

\newcommand\orcidicon[1]{\href{https://orcid.org/#1}{\mbox{\scalerel*{
\begin{tikzpicture}[yscale=-1,transform shape]
\pic{orcidlogo};
\end{tikzpicture}
}{|}}}}

%% file: Abstract.tex
%!TEX root=bare_jrnl_compsoc.tex

\begin{abstract}
While technical debt grows in absolute numbers as software systems evolve over time, the density of technical debt (technical debt divided by lines of code) is reduced in some cases.
This can be explained by either the application of refactorings or the development of new artifacts with limited Technical Debt. In this paper we explore the second explanation, by investigating the relation between the amount of Technical Debt in new code and the evolution of Technical Debt in the system. To this end, we compare the Technical Debt Density of new code with existing code, and we investigate which of the three major types of code changes (additions, deletions and modifications) is primarily responsible for changes in the evolution of Technical Debt density. Furthermore, we study whether there is a relation between code quality practices and the 'cleanness' of new code. To obtain the required data, we have performed a large-scale case study on twenty-seven open-source software projects by the Apache Software Foundation, analyzing 66,661 classes and 56,890 commits. The results suggest that writing ``clean'' (or at least ``cleaner'') new code can be an efficient strategy for reducing Technical Debt Density, and thus preventing software decay over time. The findings also suggest that projects adopting an explicit policy for quality improvement, e.g. through discussions on code quality in board meetings, are associated with a higher frequency of cleaner new code commits.  Therefore, we champion the establishment of processes that monitor the density of Technical Debt of new code to control the accumulation of Technical Debt in a software system.
\end{abstract}

%% file: Introduction.tex
%!TEX root=bare_jrnl_compsoc.tex

\IEEEraisesectionheading{\section{Introduction}\label{sec:introduction}}

\IEEEPARstart
{T}{echnical Debt} is a metaphor that captures in monetary terms, the cost of additional maintenance effort caused by technical shortcuts taken usually for expediency \cite{avgeriou2016managing}. As observed in practice \cite{digkas2017evolution}, for the majority of software systems, the amount of \td increases along evolution, due to growing size and/or reduced quality; this is aligned with software evolution laws \cite{lehman1997metrics}. However, the \emph{density} of \td, i.e., the normalized amount of \TD per line of code, remains in some cases stable, or is even reduced over time; we have observed this in our previous work on the evolution of the Apache ecosystem \cite{digkas2017evolution}. This raises the question how such systems manage to maintain or improve their Technical Debt density. 

% as it implies that certain quality aspects also improve.

There are two possible explanations for this phenomenon. The first is that these systems follow a process of systematic perfective maintenance, mostly through the application of refactorings \cite{murphy2011we}. Code refactoring is the most popular strategy for \td repayment \cite{li2015systematic}. However,  refactoring activities are rarely applied systematically in practice \cite{chatzigeorgiou2014investigating, peters2012evaluating, arcoverde2011understanding, tufano2017and, tufano2016empirical}. The second explanation is the development of new software artifacts at a level of quality that is above the average, following the clean code paradigm \cite{martin2009clean}.
Introducing new code whose Technical Debt density is kept below the system average, is not trivial and often implies the adoption of an explicit policy for \dquote{clean commits}. The sheer frequency of new commits in large or ultra-large codebases, which in the case of Google's source code can reach 16,000 on a typical workday \cite{potvin2016google}, renders this strategy even more challenging.
Clean code has recently emerged as a promising strategy facilitated by the use of Quality Gates in Continuous Integration systems.

While refactoring has been intensively studied as a technical debt repayment strategy, clean code has not. We argue that if writing clean new code is efficient in managing \td in the long term, this has important implications for both practitioners and researchers. The former can scope training activities and software development processes so as to incorporate the best possible practices for writing clean code (e.g., develop quality gates based on the \td introduced by new commits). The latter can focus their research efforts on \td prevention and repayment within new code chunks. 

In this study, we first observe how \squote{clean} is the new code compared to the \TDd of existing code for \TSEsyscountNumber projects from the Apache Software Foundation. Next, we compare the contribution of new, deleted and modified code to the changes in the system's \TDd. This allows to understand which activity affects technical debt density the most: writing new code, deleting or modifying code. Finally, we investigate if a decreasing trend in the evolution of {\TDd} is associated with the adoption of relevant practices at project management level. To this end we study whether projects that exhibit a high frequency of cleaner new code: a) provide clear guidelines to committers, so as to guarantee the quality of the newly committed code; and b) consider code quality as an important topic in project management.
%To this end we study whether projects that exhibit a high frequency of cleaner new code utilize Quality Gates, Quality Assurance tools or explicit recommendations for committing code as a means of sustaining project quality levels}.

We rely on the notion of \textit{\TDd} since absolute measures of \td (such as the number of identified violations or estimates of the effort to eliminate these violations) increase monotonically with the addition of code (i.e., absolute measures can decrease only by code deletions). \TDd is obtained as the ratio of \td effort to remediate the issues identified in a piece of code, over the corresponding lines of code. In this way, a lower \TDd of newly added code might result in a reduction of the system's total \TDd, even if the \td in absolute terms has increased. 

% \begin{figure}[!t]
%     \centering
%     \includegraphics[width=\linewidth]{figures/introfigure.png}
%     \caption{intro figure}
%     \label{fig:methodEvolution}
% \end{figure}

In terms of scope, we focus on code \td, which is the most studied type of \td in the literature  \cite{alves2016identification}, and the most important type of \td in industry  \cite{ampatzoglou2016perception}. In particular, we consider the TD incurred by code smells in the source code. In terms of granularity, we work at the method level: we monitor the introduction of new methods, and the removal or modification of existing methods. This helps to avoid incorrect classification of code changes during code evolution: changes at the instruction level can become cumbersome to track, as modification, removal and introduction of individual instructions can occur simultaneously. This is further justified in Section \ref{subsec:calculationOfNewCodeTD}. Finally, we emphasize that our scope is open source software, as they provide a long history of commit activity thereby enabling the evolutionary analysis of the study.

The rest of the paper is organized as follows: 
in Section \ref{sec:relatedwork} we present related work, i.e., studies that deal with the evolution of software quality and \td in particular, empirical studies that provide evidence on the impact and frequency of refactorings as well as recent work on the development of clean new code through the concept of quality gates.  
In Section \ref{sec:caseStudyDesign}, we present the investigated research questions, the case study design and we discuss how we monitor the contribution of new, deleted and modified code to the system's \TDd.
The results of the study on 27 Apache projects are presented in Section \ref{sec:results}, while in Section \ref{sec:discussion}, we discuss the findings, by providing interpretations and implications for researchers and practitioners. Finally, 
in Section \ref{sec:threatstovalidity} we evaluate the validity of the study, whereas 
Section \ref{sec:conclusion} concludes the paper.

%% file: RelatedWork.tex
\section{Related Work}\label{sec:relatedwork}
The current study explores the contribution that new code can have on technical debt density, as a complementary approach to applying refactoring. Therefore, this section discusses previous work on: (a) the evolution of code smells and \TD in particular, (b) evidence on the frequency and impact of refactoring, and (c) the concept of Quality Gates that focus on ensuring a desired level of quality in new commits.

\subsection{Evolution of Code Smells}\label{subsec:evolutionOfSoftwareQuality}
Lehman's seventh law of software evolution states that \textit{the quality of a system will appear to be declining during its evolution, unless proactive measures are taken} \cite{lehman1996laws}. To this end, many studies have explored the evolution of code quality, and if indeed this law stands in practice. Since this paper focuses on code \TD, we scope this sub-section to the evolution of code smells.

One of the first studies that investigate the evolution of code smells was conducted by Olbrich et al. \cite{olbrich2009evolution}. On their study, they investigate the evolution of two code smells, God Class and Shotgun Surgery, on two projects by the \asf, namely Apache Lucene and Apache Xerces. The results of their study, show that during the software development, there are phases where the number of those code smells can either increase or decrease and those phases are not affected by the size of the systems.

Chatzigeorgiou and Manakos \cite{chatzigeorgiou2014investigating} have also investigated the evolution of code smells in \os object-oriented projects. They used historical data of two \oss projects, namely: JFlex and JFreeChart and studied the evolution of four code smells namely: Long Method, Feature Envy, State Checking, and God Class smells. The results of their study show that as the projects evolve over time the number of code smells tends to increase, which confirms the Lehman's seventh law. Furthermore, they have also found evidence that developers rarely perform targeted refactoring activities to remove smells. In most of the cases, if code smells disappear over time, this is a side effect of regular maintenance (e.g. removal of code). Another interesting finding was that a significant percentage of smells was not the results of software ageing, but smells were present right from the first version of the code in which they reside.

Tufano et al. \cite{tufano2017and} also studied the evolution of code smells with the goal of understanding when and why code smells are introduced into the projects and observe their life cycle. 
The study was based on five code smells: Blob Class, Class Data Should be Private, Complex Class, Functional Decomposition, Spaghetti Code. The results indicate that: (a) in the majority of the cases the code smells are introduced into the projects with the creation of the corresponding classes or files, (b) while projects evolve over time, \dquote{smelly} code artifacts tend to become more problematic, (c) new code smells are introduced when software engineers implement new features or when they extend the functionality of the existing ones, (d) the developers who introduce new code smells into the projects, are the ones who work under pressure and not necessarily the newcomers, and (e) the majority of the smells are not removed during the project's evolution and few of them are removed as a direct consequence of refactoring operations.

Peters and Zaidman \cite{peters2012evaluating} studied the lifespans of the following code smells: God Class, Feature Envy, Data Class, Message Chain Class, and Long Parameter List. They developed a tool called SACSEA and used it to analyze the history of eight \oss projects. Their findings show that while projects evolve, the number of code smells increases. Furthermore, they have also found that although developers are aware of the existence of the code smells they do not perform refactorings. Finally, their findings imply that `simpler code smells (e.g. Feature Envy Methods) are refactored more often, without any evidence on whether this happens intentionally or not.

Digkas et al. \cite{digkas2017evolution} analyzed and tracked the evolution of \td of \ECSAsyscount \os Java projects by the \asf, over a period of 5 years. In order to track and detect issues that incur \td they relied on \sq. The results of their study show that on the one hand, there is a significant increasing trend on the size, complexity, number of \TDIs, and the total \TD over time, which seems to confirm the software aging phenomenon. But on the other hand, when \td is normalized over the non-commented lines of code of the project, an evident decreasing trend over time is present for many of the projects. This could possibly be attributed to: (a) developers that perform refactoring activities and fix some of the open \TDIs; or (b) developers that introduce better quality code in each commit (compared to the project's existing code base).

Prior research provides evidence that the number of code smells increases over time \cite{digkas2017evolution}, \cite{peters2012evaluating} and that smells are often introduced along with the creation of the corresponding classes/files \cite{chatzigeorgiou2014investigating}, \cite{tufano2017and}. However, these studies have not investigated the association between overall trends in system quality with the cleanness of new code or the quality practices followed in a project so as to provide insight into the potential of clean new code as a means of reducing \TD.

% Architectural decayis caused by repeated changes to a system during its lifespan
% conducting an empirical study of changes found in softwarearchitectures spanning several hundred versions of 14 open-source systems
% findings regardingthe frequency of architectural changes in software systems
% a system’s versioning scheme is not anaccurate indicator of architectural change: major architecturalchanges may happen between minor system versions
% a system’s architecturemay be relatively unstable in the run-up to a release

\subsection{Refactoring Frequency and Impact}\label{subsec:refactoring}
In this sub-section, we first discuss the frequency at which refactorings are applied, and then we provide evidence on the impact of refactorings on code quality.

Evidence shows that developers rarely apply code refactorings to remove smells. Arcoverde et al. \cite{arcoverde2011understanding} studied the lifespan of code smells within software projects and investigated why developers tend to perform very few refactorings. The results of their explanatory survey show that developers are reluctant to perform refactorings in order to avoid API modifications.

Yamashita and Moonen \cite{yamashita2013developers} also tried to shed light on why the developers do not perform refactorings on their projects. They conducted an exploratory survey with 85 developers to investigate how familiar they are with the notion of code smells. The results show that one third of the interviewed developers are not aware of code smells or have limited knowledge about about them. Furthermore, many of them expressed the lack of good supporting tools that would help them identify smelly pieces of code as candidates for refactoring.  
% Yamashita and Moonen \cite{yamashita2012code} have also conducted another study that shows that the maintainability of a file is effected by the number of code smells. As the number of code smells grows the file's `maintainability index' decreases.

Murphy-Hill et al. \cite{murphy2011we} studied broader developers' refactoring habits. Similar to other studies they found that the developers rarely perform refactoring activities and usually, when they do, they combine those refactorings with other code changes. Finally, they observed that even when developers do perform refactoring activities, they do not systematically record them, e.g. as a message on their commits.

A Google initiative in 2009 asked engineers to participate in a companywide \dquote{Fixit} week, focusing on resolving warnings issues by a static analysis tool. Only 16\% of the total number of warnings were actually fixed, despite the fact that almost half of the reviewed issues resulted in filing a bug report \cite{sadowski2018lessons}. It is also noteworthy that Google developers deemed 74\% of the issues raised early (i.e. at compile time) as \squote{real problems}, compared to 21\% of suggested changes for already checked-in code.

% \TD can be found in many components of a software project (i.e. architecture, source code, comments, etc). Maldonado et al. \cite{maldonado2017empirical} have conducted an empirical study on the removal of Self-Admitted Technical Debt (SATD). SATD can be found on the source code comments that indicate Technical Debt. The results of their study show that the removal of the SATD items could take from several months and up to a decade and usually the developer who introduced them is also the one who removes, during a bug fixing or a feature implementation.

A number of studies have empirically investigated the effect of refactoring application on various software qualities. Stroggylos and Spinellis \cite{stroggylos2007refactoring} examined the logs in the version control systems of four \oss projects to extract the commits where refactorings had been performed. Next, they measured the effect of refactorings on selected software metrics. The findings reveal that, despite the expectation that refactorings would improve  software quality, measurements on the examined systems indicate the opposite. In particular, it was found that refactoring caused a non trivial increase in metrics related to cohesion and coupling.

% Kannangara et al. \cite{kannangara2013impact} assessed the impact of ten refactoring techniques on code quality. Their findings show that ``Replace Conditional with Polymorphism'' had the highest percentage of improvement in the software's quality and the ``Introduce Null Object'' had the lowest. Furthermore, they have also found that the refactorings effect positively the analyzability of the software.

To investigate how specific quality factors are affected by refactoring, Bois and Mens \cite{du2003describing} proposed a  formalism based on abstract syntax tree representation of source code and projected the impact of refactoring on internal quality metric values defined on this representation. The selected refactorings were Extract Method, Encapsulate Field, and Pull Up Method. Although the study is not focused on obtaining extensive empirical results, the application of the examined refactorings can have a mixed effect on different metrics (such as size, coupling and cohesion ones). 

% Kataoka et al. \cite{kataoka2002quantitative} proposed a quantitative evaluation method to measure how the maintainability index, changes after a refactoring activity is performed. In order to evaluate if a refactoring had a positive or negative impact, they compared the coupling before and after the refactoring performed.

% Liu et al. \cite{liu2013monitor} proposed a monitor-based refactoring framework which helps the developers to fix the code smells on their project more frequently, faster, and reduce their lifespan. Their results show that the resolution of the smells increased significantly, the ratio of introduced smells has decreased significantly, and also the lifespan of the code smells - which means that it helps the developer to find the smells and fix them.

Wilking et al. \cite{wilking2007empirical} conducted a controlled experiment to investigate how refactorings affect the maintainability and modifiability of the projects. Their approach consisted in randomly inserting 15 syntactical and 10 non-syntactical errors into code and they measured the time that is needed to fix them. Concerning the effect of the refactorings on the  modifiability, they evaluated it by adding new implementation requirements and they measured the time and the Lines of Code that are required in order to implement them. The results of their controlled experiment show that there is no direct effect of improved maintainability or modifiability due to refactoring.

In another study, Alshayeb concluded that refactoring application does not necessarily improve external quality attributes such as adaptability, maintainability, understandability, reusability and testability \cite{alshayeb2009empirical}. By applying refactoring techniques as defined by Fowler \cite{fowler2018refactoring} on three software systems and measuring the impact on selected software metrics, an immense variation of the refactoring effect was found. Thus, the author concluded that he was unable to validate that refactoring as a practice improves quality.  

A multi-project study on 23 \oss projects and more than 29000 refactoring operations to study the effect on internal quality attributes was reported by Chavez et al \cite{chavez2017does}. The analysis revealed that 65\% of the refactoring operations improve the internal quality as measured by a wide set of metrics, while 35\% of the refactorings keep the quality attributes unaffected. 

Although the above set of research studies is not exhaustive, most of the findings agree on the limited adoption of refactorings in practice and a rather mixed effect on software qualities, at least for quality aspects that can be captured by source code metrics. Such evidence calls for the systematic study of other strategies to sustain or improve quality in software systems over time.  

\subsection{Quality Gates}\label{subsec:qualitygates}
The aforementioned law of declining software quality during software evolution entails that it is not sufficient to write good code in the first place; code has to be \textit{kept clean over time}. As Martin vividly states, this practice adheres to the \dquote{Boys Scouts of America} rule to \textit{leave the campground cleaner than you found it} \cite{martin2009clean}. The simple and rational strategy of checking-in code that is cleaner than the average of the existing code-base will eventually yield continuous improvement in software quality. In this sub-section we focus on this strategy for reducing \TDd, i.e., by ensuring that new code commits do not violate a particular set of rules (i.e. do not introduce new \TDIs) \cite{martin2009clean,lehman1996laws}. This strategy is based on the notion of quality gates \cite{falessi2017if}. 

Software engineers can use quality gates in order to set constraints, i.e., reject commits that contain any or particular code or design inefficiencies: In case a \squote{zero-defect} policy is adopted, the new code will essentially be TD-free. In practice, quality gates can be more flexible i.e., reject commits that contain smells of a given severity, type or priority level. Quality gates can be easily combined with Continuous Integration (CI) practices setting the maximum level of \TD that is acceptable for new commits to the project’s repository.

Janus et al. in 2012 \cite{janus20123c} have proposed the 3C Approach. It is an extension to the Agile Practice Continuous Integration and it relies on quality gates for agile quality assurance combining software metrics with Continuous Integration. The proposed automated metric-based Quality Gate checks the source code and ensures that it does not exceed any of the defined thresholds before committing it to the version control system. This way the internal Software Quality is assured. In order to deploy and validate their method, they analyzed an agile project that was developed by a German Automotive Industry company and the results show that a significant improvement of its internal quality can be achieved.

Suryanarayana et al. \cite{suryanarayana2015software} argued that smells are the result of violating some of the best practices and indicate higher-level design problems. They classified the smells based on the primary object-oriented design principle that they violate, namely: abstraction, modularization, and hierarchyduplicate abstraction, insufficient modularization, and multipath hierarchy smells. Based on an experiment/study that they conducted the found that one of the reasons that code smells are inserted into the project is the time pressure, thus the developers prefer to perform a quick (and dirty) fix rather than an appropriate solution. Finally, in order to avoid this symptom, they proposed a design quality gate process that checks if the modified/inserted code violates any of the predefined design-level rules. 
% One of the most important objectives of Agile methods, is increasing the speed of deploying new releases to production. On the one hand, if the software released without any qualification, there is a high risk that the release could be faulty. On the other hand, software qualification is a time-consuming process and slows down the release speed / time to market.

Schermann et al. \cite{schermann2016towards} acknowledge that Quality gates, as steps that ensure the reliability of code changes, lead to an inherent trade-off between sustaining a fast pace and risking a lower release quality. To address this issue they
proposed a model where software releases are evaluated based on the Confidence (reliability) and Velocity (publishing speed). Their Confidence-Velocity Categorization Model consists of the following four categories: Cautious (low Velocity and high Confidence), Balanced (high Velocity and high Confidence), Problematic (low Velocity and low Confidence), and Madness (high Velocity and low Confidence).

Nevertheless, according to the empirical investigation by Vassallo et al. \cite{vassallo2018continuous} Continuous Code Quality (CCQ) is not applied in practice. The authors attribute the low use of CCQ to the fact that code quality is not always the top priority for development teams but also the unawareness of how to properly set up quality gates.  

% \begin{itemize}
%     \item Cautious (Velocity: low, Confidence: high), where the companies are really careful before the release of a new version
%     \item Balanced (Velocity: high, Confidence: high), This category is defined by high velocity, thus taking advantage of reduced time to market and early customer feedback, combined with high confidence about the quality of releases
%     \item Problematic (Velocity: low, Confidence: low), Companies in this category lack confidence in their quality gates. Reasons include, amongst others, missing regression testing, the absence of or not enough code reviews, insufficiently maintained test suites (both automated and manual)or their wrong execution, a shortage in quality assurance personnel, or unclear roles and responsibilities regarding quality assurance
%     \item Madness (Velocity: high, Confidence: low), Madness is the combination of problematic with high velocity. Companies in this category benefit from short release cycles, thus early customer feedback and reduced time to market. \TDIs might be fixed fast, but the lack of proper quality gates make releases risky and stressful
% \end{itemize}

%% file: CaseStudyDesign.tex
%!TEX root=bare_jrnl_compsoc.tex

\section{Case Study Design}\label{sec:caseStudyDesign}

Case study is an  empirical method that is used for studying phenomena (e.g., projects or activities) in a real-life context \cite{wohlin2012experimentation}. The case study of this paper has been designed and is presented according to the guidelines of Runeson et al. \cite{runeson2012case}.

\subsection{Goal and Research Question}\label{subsec:proposedmethodology}
The goal of this study is to compare addition, deletion and modification of code regarding their impact on \TDd. Moreover, to provide further insight to the relevant strategies, we study whether code quality practices are associated with the cleanness of new code. Therefore, we formulate two relevant research questions.

\noindent
\textbf{RQ$_{1}$:} \rqone

RQ$_{1}$ aims at investigating whether changes in technical debt density from one code revision to the next are primarily associated with addition of new code, deletion or modifications of existing code. Each type of change can incur a negative or positive effect on the system's \td density depending on the quality of the code that is added, modified or removed. Code modifications can sometimes be related to the application of refactorings, but in the general case we assume that code changes are the result of maintenance and not necessarily targeting the removal of inefficiencies.

\noindent
\textbf{RQ$_{2}$:} \rqtwo

RQ$_{2}$ aims at investigating whether a high percentage of cleaner new code commits is related to the use  of practices targeting code quality. In terms of relevant practices we study two project management aspects: (a) the existence of commit guidelines (i.e. what the developer should have in mind before committing his/her code) which are directly or indirectly related to the avoidance of TD rule violations; and (b) the extent to which quality related issues (e.g., code improvement, code quality, refactorings, etc.) are discussed in project board meetings. Assessing the strength of the association can lead to interesting actionable outcomes, which can guide project managers on how to control the quality of their projects.
% USED TO BE
%\hl{RQ$_{3}$ aims at investigating whether a high percentage of cleaner new code commits can be attributed to the use of Quality Gates or Quality Assurance tools in the development process or to the existence of explicit recommendations on how to write good quality code.} \textcolor{red}{We study the XXXXXX in two sub-questions, namely XXXXXXXXX} 

\subsection{Cases and Units of Analysis}\label{subsec:casesunitsanalysis}
This study is characterized as multiple, embedded case study\cite{runeson2012case}, in which the cases are \oss (OSS) projects and the units of analysis are the revisions across the project history; we analyse changes to the  system's \TDd in these revisions. The reason for selecting to perform this study on OSS systems is the vast amount of data that is available in OSS repositories, in terms of revisions and classes, as well as quality-related practices. The long history that is available for each OSS project enables researchers to observe overall trends in the evolution of their quality. To retrieve data from only high-quality projects that evolve over a period of time, we have selected to investigate the projects presented in Table \ref{tab:projectDetails}. We have decided to focus on Apache projects (similarly to the studies by Tan et al.  \cite{tan2018towards} and Tufano et al. \cite{tufano2017and}) since the Apache Software Foundation, as an OSS development organization, has a reputation for high quality projects, for putting emphasis on process and quality improvement as well as for long-lasting projects.

The project selection process was based on the following criteria:
\begin{enumerate}[label=\alph*.]
    \item The project should be active (based on the date of its last commit) and therefore still maintained. This criterion aims at ensuring that the analyzed projects are still undergoing development and thus the studied practices are up-to-date. A similar prerequisite has been set by Rausch et al. \cite{rausch2017empirical} who studied the build failures in Continuous Integration (CI) workflows of \oss.
    \item The software should be written in Java and use Maven as a build tool. This ensures that the project can be built and allows the retrieval of the project's language version from the corresponding \pom file. 
    \item The software should contain more than \TSEclasses classes to ensure the inclusion of systems with a substantial size, functionality and complexity. A minimum number of system classes has also been set as a project selection criterion in the studies by Tan et al. \cite{tan2018towards} and Olbrich et al. \cite{olbrich2009evolution}.
    \item The software should have more than \TSEcommits commits and should be under development for at least \TSEyears years. We have included this criterion for similar reasons to the previous criterion and to be able to observe trends in the evolution of the projects quality. Moreover, this number of revisions provides an adequate set of repeated measures as input to the statistical analysis. A minimum number of commits has also been used as a criterion in other studies on software evolution \cite{rausch2017empirical}, \cite{tan2018towards}, \cite{olbrich2009evolution}, \cite{peters2012evaluating}, \cite{herraiz2006comparison}.   
\end{enumerate}

The selection of Apache projects enabled us to perform the analysis on RQ$_{2}$ which is based on the availability of minutes for Apache Board Meetings. However, we should note that there are also other Apache projects fulfilling the above mentioned criteria beyond those included in our dataset. Due to the complicated data analysis we have excluded projects that are extremely large, either in the number of  classes or the number of commits.

% ////////////////////////////////
% We have used similar criteria on our previous studies \cite{digkas2017evolution,digkas2018developers}. 
% Namely on \cite{digkas2017evolution} we chose projects from \asf that their main programming language is Java, and they had at least two years of evolution/development.
% On \cite{digkas2018developers} we also studied projects by the \asf that 1) their main language was Java, 2) had at least 100 classes, 3) had at least 1000 commits and 4) there was commit activity the year that we conducted the study.

% ////////////////////////////////

\input{tables/CaseStudySelectedProjects.tex}

\subsection{Tracking the Types of Changes}\label{subsec:calculationOfNewCodeTD}

Considering that software systems evolve through a number of revisions and that in each revision several types of changes may occur simultaneously, we look at the three major types of method changes: 
% the development of new code, the deletion or the modification of existing code. 
the development of new methods, the deletion or the modification of existing ones. These primary types of evolutionary changes have been considered in other studies as well \cite{beyer2017}, \cite{jaafar2016}, \cite{xing2005analyzing} and \cite{Dagenais2011}.
As already mentioned,  monitoring changes at the instruction level would be more complex and less accurate considering that several types of changes can simultaneously occur in some statements (e.g., modification and introduction of new code). Furthermore, tracking changes at the instruction level is challenging, as one would have to map each instruction (in a particular revision) to the corresponding instruction in the previous revision. This process is complicated by the insertion of new statements, comments, blank lines, etc. Therefore, to be certain about the classification of changes, we monitor changes at the method level.

\begin{figure*}[!t]
    \centering
    \includegraphics[width=\linewidth]{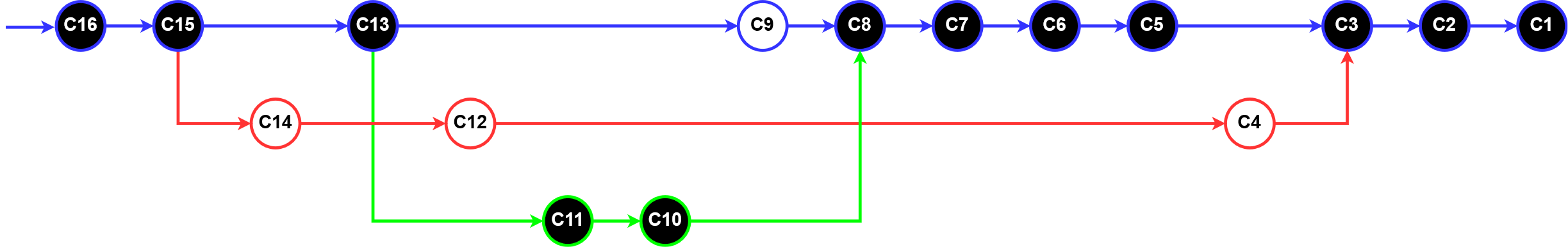}
    \caption{Seeking the longest path between commit nodes}
    \label{fig:git_revisions_diagram}
\end{figure*}

Similarly, instead of assessing the entire \td of the analyzed systems, i.e., considering violations on every individual line of code, we have opted to consider only TD that can be mapped to class methods. In other words, we consider only \sq rule violations which reside in class methods. The reason is that \TDIs which occur at the class- or file-level (e.g., \textquote{The default unnamed package should not be used}) are not associated with particular lines of code; as a result it would not be possible to assess what kind of code change caused their introduction or removal. 

At each revision a method can be added, deleted, modified or remain unchanged. 
According to the stated research questions, the goal of this study is two-fold: (a) since multiple types of changes might occur simultaneously, to identify the type of change that has the largest impact on \TDd change, i.e. whether the \TDd change from one revision to the next is mostly due to the modification, the deletion or the addition of new methods, and (b) to investigate whether the frequency of \squote{clean} new code commits is related to overall project policies.

\subsection{Data Collection}\label{subsec:dataCollection}

To analyze the projects and measure \TD throughout their evolution, we have used \sq. \sq relies on a set of rules which are checked by static source code analysis; every time a piece of code breaks one of those coding or design rules, a \td issue is raised. Thus, \sq estimates the effort (in minutes) required to eliminate the identified \TD issues\footnote{In this study we have considered \TD issues reported as code smells by \sq}. This effort is obtained by assigning a time estimate for fixing each type of problem and by multiplying all issues of the same type with that estimate. It should be noted that the Apache Foundation ecosystem, has a dedicated \sq instance for quality control in its projects. Currently, 336 Apache projects are continuously monitored through SonarCloud, and 90.1{\%} pass the quality criteria set by the development teams\footnote{\url{https://sonarcloud.io/organizations/apache/projects}}. \sq reports various types of problems, namely code smells, bugs (issues representing something wrong in the code), vulnerabilities, code duplications and lack of test coverage. We note that in this study, we only consider code smells, since the other two types of problems (i.e., bugs and security vulnerabilities) do not fit the definition of TD \cite{avgeriou2016managing}, in the sense that they do not concern maintainability or evolvability.

For RQ$_{1}$ we measure the contribution to the \TDd change at each revision that is due to (a) new methods, (b) removed methods, and (c) modified methods. For RQ$_{2}$ we consider the revisions in which the \TDd of new code is lower than that of existing code.  

\subsubsection{\textbf{Data Collection for Answering RQ$_{1}$}}
We have devised a process for analyzing git repositories which can be outlined in the following phases and individual steps:

\noindent
\textit{Phase 1: Retrieval of commits}

\begin{enumerate}
    \item First, the Git history for the project under study is retrieved from its master (default) branch since it reflects the production-ready state of the project.
    \item All commits are sorted to form a time series of revisions that have been performed on the source code. This process is non-trivial since even on the master (default) branch a commit can have multiple parents. To treat the data as a single time series, a single parent should be chosen for each commit without forming any branches. For the cases of commits with more than one parent, we have employed an algorithm aiming at identifying the longest path between the commit node under examination and the start node (i.e. the only node with no parent). As an example, Fig. \ref{fig:git_revisions_diagram} shows that commits 13 and 15 have more than one parents. We select for analysis the path consisting of the black nodes since it forms the longest possible path. Had we selected the series of commits as formed chronologically we would have run into inconsistencies among revisions: for example, if we analyze CM1, CM2, CM3, CM4, CM5 and so forth, any change on \TD at commit CM4 would not be valid for the chronologically subsequent commit CM5, and changes to TD across revisions would yield irrational results. At the same time, the longest path yields the largest number of commits to be analyzed and thus results in a higher granularity for the analysis. 
    \item To reduce the computation time, a filtering step is applied: we ignore transitions between successive commits that do not involve any changes to Java files. We do this because the analysis of multiple revisions of large projects in \sq is extremely  computationally-intensive resulting in several hours or even days for analyzing the entire history of the selected \os projects.  
    \item From all commits submitted for analysis to \sq we retain only the successfully analyzed commits. The reason is that several commits may fail to analyze for various reasons, such as an incorrect \pom file that prohibits the build of the project. 
\end{enumerate}

\noindent
\textit{Phase 2: Mapping of \TDIs to methods}

To map the identified \TDIs to the class methods of each revision we perform the following steps:

\begin{enumerate}
    \item First, for each revision, we retrieve all \TDIs by performing the corresponding query to the \sq database. 
    \item Next, we map the identified \TDIs to the methods of the corresponding revision. This is performed by matching the line in which each \TDI is reported by \sq (in case the \TDI concerns multiple lines, \sq reports the first one) with the method containing that line.
\end{enumerate}

\noindent
\textit{Phase 3: Tracking method changes}

In order to associate variations in the overall \TDd of a system with code changes at the method level, we track the type of change (introduction, deletion, modification) occurring to each method as follows:

\begin{enumerate}
    \item For the new and deleted files of each revision (obtained from git history) we obtain their representation in the form of an Abstract Syntax Tree (AST)\footnote{The AST is obtained through the Eclipse Java Development Tools (JDT)}. For each new/deleted file, we extract all its methods from the AST representation and then tag all these methods as new/deleted, respectively.   
    \item For the modified files of each revision we track new/deleted/modified/unchanged methods in each transition with the help of the \href{https://github.com/SpoonLabs/gumtree-spoon-ast-diff}{Gumtree Spoon AST Diff tool} \cite{falleri2014fine}.
\end{enumerate}

\textit{Phase 4: Calculating the contribution of new/deleted/modified methods to the change in the system's \TDd}

Finally, we need to calculate, for each revision in the system's history, the contribution of new/deleted/modified methods to the change of the system's \TDd. Let us consider a transition from revision \textit{t-1} to revision \textit{t}. To segregate the contribution of each type of change and at the same time ensure that the sum of all contributions is equal to the total change in the system's \TDd, we subtract the \TDd of the previous revision from the \TDd that is derived by the addition, removal or modification of code. The calculation is outlined in the following formulas:

% \begin{enumerate}
%     \item TD Density of newly added methods: 
%     \[\frac{\text{Amount of introduced TD}}{\text{LOC of new method}}\]
    
%     \item TD Density of deleted methods:
%     \[\frac{\text{Amount of removed TD}}{\text{LOC of deleted method}}\]
    
%     \item TD Density of modified lines in updated methods:
%     \[\frac{\text{Change in the amount of TD for the modified methods}}{\text{Change in the LOC of modified method}}\]
    
%     \item TD Density of existing methods:
%     \[\frac{\text{TD of existing methods}}{\text{LOC of existing methods}}\]
% \end{enumerate}

\textbf{Contribution of new methods}

$\Delta TD_{density}(new)$ = 
\begin{equation}\label{eq:TDdNEW}
\frac{\text{TD$_{t-1}$ + TD$_{new(t)}$}}{\text{LOC$_{t-1}$ + LOC$_{new(t)}$}} - TD_{density}(t-1)
\end{equation}

\textbf{Contribution of deleted methods}

$\Delta TD_{density}(deleted)$ =
\begin{equation}\label{eq:TDdNEW}
\frac{\text{TD$_{t-1}$ - TD$_{deleted(t)}$}}{\text{LOC$_{t-1}$ - LOC$_{deleted(t)}$}} - TD_{density}(t-1)
\end{equation}

\textbf{Contribution of modified methods\footnote{In eqs. (2), (3) the denominator can not obtain the value zero under real circumstances, as this would imply that all lines of code are deleted or modified in a certain revision}}

$\Delta TD_{density}(modified)$ =
\begin{equation}\label{eq:TDdMODIFIED}
\frac{\text{TD$_{t-1}$ $\pm$ $\Delta$TD$_{modified(t)}$}}{\text{LOC$_{t-1}$ $\pm$ $\Delta$LOC$_{modified(t)}$}} - TD_{density}(t-1)
\end{equation}

As a result, the change in the system's \TDd is equal to the sum of the individual contributions:

\begin{align}
\begin{split}
\Delta TD_{density}(system) &= \Delta TD_{density}(new) \\
&+ \Delta TD_{density}(deleted) \\
&+ \Delta TD_{density}(modified)
\end{split}
\end{align}

It should be noted that the data collection process has led to an enormous data set of approximately 1.4TB. 
A \href{https://drive.google.com/drive/folders/1mxher2vkE68GzKAkz1Y7rjVB0_hTihzt}{replication package} with all data required to study the two RQs is available online\footnote{Replication package is available at \url{https://drive.google.com/drive/folders/1mxher2vkE68GzKAkz1Y7rjVB0_hTihzt}}.

\subsubsection{\textbf{Data Collection for Answering RQ$_{2}$}}
To explore the association of code quality practices and the quality of new code (RQ$_{2}$), we use the results of the descriptive statistics (the percentage of commits in which the new code is cleaner compared to existing code [CLEAN\_CODE FREQ]), and two other variables. The first variable [COMMIT\_GUIDELINES] is binary, and is set to true if: (a) the website of the project has clear and public guidelines for committers (usually termed \dquote{How to ...}); and (b) if at least one of the guidelines is not a purely aesthetic/formatting guideline (e.g., indent your code using tabs) and is directly or indirectly related to the rules being checked by \sq. A detailed reporting of how each project has been evaluated with respect to Commit Guidelines is presented in the online replication package. The second variable [PROJECT\_BOARD\_MEETINGS] is related to the emphasis of the project board on quality issues. In particular, for each Apache Software Foundation project, there is a regular meeting (usually every 3 months), in which the  managers or key contributors of the project discuss the open issues and strategies for further improvement. To assign a value to [Project Board Meetings] variable, we have parsed the minutes of these meetings, aiming to identify discussions related to:

\begin{itemize}
    \item \textit{quality control (QC)}, for which we searched for the keywords: \dquote{software quality}, \dquote{code quality}, \dquote{code improvement}, \dquote{code review}, \dquote{guideline},  or \dquote{sonar}
    \item \textit{refactorings (REF)}, for which we searched for the occurrence of \dquote{refactoring} and \dquote{clean up}
\end{itemize}

\noindent
and recorded the number of meetings in which each term was identified (variables [QC] and [REF]). Next, we calculated and rounded the MEDIAN value for the two variables ([QC] and [REF]). Every value that was higher than the rounded median was characterized as HIGH, whereas the rest as LOW. Projects characterized as HIGH in both perspectives, have been marked as HIGH in the [Project Board Meetings] variable, whereas all the rest as LOW. In other words, we classify projects in two categories based on the frequency by which project board meetings deal with code quality and refactoring strategies.

\subsection{Data Analysis}\label{subsec:dataAnalysis}
To answer the research questions using the collected data we carried out both descriptive and inferential statistical analysis as follows:

\noindent

\subsubsection{\textbf{Data Analysis for Answering RQ$_{1}$}}
The investigation of the contribution of each type of code change to the variation of the system's \TDd across revisions is quite complicated, as the effect of the three types of changes (additions, deletions and modifications of methods) has to be taken into account. In particular, we are interested in observing whether positive (negative) changes in the system \TDd co-exist with positive (negative) contributions stemming from new/deleted/modified methods. 

\subsubsection*{Independent Variables}
Contribution to \TDd of new, deleted and modified code. These are categorical variables.

\noindent
Categories: Leading to a decrease, increase or no change (stable) in \TDd.

\subsubsection*{Dependent Variable}
Direction of \TDd change during a transition from one revision to the next (the cases when the \TDd remained stable are rare and are omitted for clarity). This is a categorical variable.

\noindent
Categories: Increase/Decrease.

\subsubsection*{Analysis}
We first obtained contingency tables to describe the relationship between the two categorical variables. It should be noted that to investigate the effect of each type of change, we retained only the transitions when two types of changes occurred simultaneously, that is when two types of changes \textit{compete} for the effect on the overall change in \TDd. In case only one type of change has occurred during a transition, then it is obvious that the change in \TDd will be the result of this single change, and thus including such transitions in the data set would lead to misleading results. The results are displayed in the form of heat maps. For each project (row) three individual heat maps are shown, one for the contribution of new, deleted and modified code, respectively. 

To further investigate this relationship, we performed a chi-squared test between the two categorical variables, to determine whether there is a significant relationship between them. 

\noindent
\fbox{\parbox{\columnwidth}{Null Hypothesis H$_{0}$: assumes that that there is no relationship between the direction of change in the system's \TDd (decrease or increase) and the corresponding direction of change caused by new, deleted or modified methods.

Alternative Hypothesis H$_{1}$: Assumes that there is an association between the two variables.}
}

Finally, to shed light into the effect of new methods vs. the effect of modified methods on code improvement, we illustrate graphically (in a bar chart) the percentage of transitions in which a reduction in the system's \TDd co-occurred with positive contributions (i.e. leading to a decrease of \TDd) by new and modified methods, respectively.

\subsubsection{\textbf{Data Analysis for Answering RQ$_{2}$}}
To answer RQ$_{2}$, we explored whether: (a) projects that provide commit guidelines or (b) projects in which Project Board meetings often refer to code quality, are having a statistically significant higher average number of commits of cleaner code, compared to projects that do not provide guidelines and do not discuss code quality often.

\subsubsection*{Independent Variables}
Binary variable [COMMITGUIDELINES] representing whether a project has commit guidelines related to code quality. Values: YES, NO.

\noindent
Binary variable [PROJECTBOARDMEETINGS] representing the frequency by which quality issues are discussed in project board meetings. Values: LOW, HIGH. 

\subsubsection*{Dependent Variable}
The  percentage  of  commits  in which  the  new  code  is  cleaner  compared  to  existing  code [CLEAN\_CODE FREQ].

\subsubsection*{Analysis}
We explored the discriminative power of the [COMMIT\_GUIDELINES] and the [PROJECT\_BOARD\_MEETINGS] in terms of the [CLEAN\_CODE FREQ] variable.  To this end we have used boxplots to illustrate any differences in the percentage of cleaner code commits between projects that do not provide commit guidelines vs. those that provide them, and between projects that often refer to code quality issues vs. those that do it less often. Moreover, we have performed independent samples t-test to test any statistically significant differences. being 

\noindent
\fbox{\parbox{\columnwidth}{Null Hypothesis H$_{0}$: assumes that that there is no difference in the percentage of cleaner code commits, regardless of any adopted code quality practices (means are equal).

Alternative Hypothesis H$_{1}$: Assumes that the percentage of cleaner code commits differs depending on the adopted code quality practices (means are not equal).}}

%% file: tables/CaseStudySelectedProjects.tex
%!TEX root=bare_jrnl_compsoc.tex

\begin{table}[!t]
    \caption{Selected Projects}
    \label{tab:projectDetails}
    \centering
    %\begin{tabular}{l|p{.5\linewidth}|r|r|p{.07\linewidth}}
    \begin{tabular}{l|r|r|r}
        \toprule
        \multicolumn{1}{c|}{Project} & 
        %\multicolumn{1}{c|}{Short Description} & 
        \multicolumn{1}{c|}{Classes} & \multicolumn{1}{c|}{NCLOC} & \multicolumn{1}{p{.15\linewidth}}{Analyzed Revisions}\\
        \midrule
        \href{https://github.com/apache/accumulo}{Accumulo} & 5840 & 428543 &  \multicolumn{1}{r}{2863} \\
        \href{https://github.com/apache/atlas}{Atlas} & 932 & 87637 &  \multicolumn{1}{r}{1454} \\
        \href{https://github.com/apache/beam}{Beam} & 3757 & 176663 &  \multicolumn{1}{r}{2780} \\
        \href{https://github.com/apache/calcite}{Calcite} & 2606 & 186633 &  \multicolumn{1}{r}{1448} \\
        \href{https://github.com/apache/cayenne}{Cayenne} & 2615 & 164170 &  \multicolumn{1}{r}{2116} \\
        \href{https://github.com/apache/cxf}{CXF} & 4111 & 353085 &  \multicolumn{1}{r}{5079} \\
        \href{https://github.com/apache/deltaspike}{DeltaSpike} & 951 & 46182 &  \multicolumn{1}{r}{513} \\
        \href{https://github.com/apache/drill}{Drill} & 4655 & 316552 &  \multicolumn{1}{r}{1316} \\
        \href{https://github.com/apache/incubator-dubbo}{Dubbo} & 943 & 61865 &  \multicolumn{1}{r}{728} \\
        \href{https://github.com/apache/flink}{Flink} & 5632 & 341149 &  \multicolumn{1}{r}{5329} \\
        \href{https://github.com/apache/flume}{Flume} & 790 & 51897 &  \multicolumn{1}{r}{789} \\
        \href{https://github.com/apache/giraph}{Giraph} & 1414 & 72972 &  \multicolumn{1}{r}{668} \\
        \href{https://github.com/apache/jackrabbit}{Jackrabbit} & 2883 & 273574 &  \multicolumn{1}{r}{4260} \\
        \href{https://github.com/apache/jclouds}{jclouds} & 5687 & 227459 &  \multicolumn{1}{r}{4323} \\
        \href{https://github.com/apache/knox}{Knox} & 1083 & 51429 &  \multicolumn{1}{r}{1033} \\
        \href{https://github.com/apache/kylin}{Kylin} & 1658 & 128531 &  \multicolumn{1}{r}{3205} \\
        \href{https://github.com/apache/metron}{Metron} & 1433 & 72579 &  \multicolumn{1}{r}{548} \\
        \href{https://github.com/apache/myfaces}{MyFaces} & 1843 & 174158 &  \multicolumn{1}{r}{1211} \\
        \href{https://github.com/apache/nifi}{NiFi} & 4256 & 371031 &  \multicolumn{1}{r}{1490} \\
        \href{https://github.com/apache/oozie}{oozie} & 1082 & 97597 &  \multicolumn{1}{r}{587} \\
        \href{https://github.com/apache/openwebbeans}{OpenWebBeans} & 561 & 44299 &  \multicolumn{1}{r}{1583} \\
        \href{https://github.com/apache/pdfbox}{PDFBox} & 1279 & 136916 &  \multicolumn{1}{r}{3758} \\
        \href{https://github.com/apache/pulsar}{Pulsar} & 1837 & 147182 &  \multicolumn{1}{r}{1503} \\
        \href{https://github.com/apache/sis}{SIS} & 1948 & 181588 &  \multicolumn{1}{r}{828} \\
        \href{https://github.com/apache/storm}{Storm} & 3958 & 243574 &  \multicolumn{1}{r}{738} \\
        \href{https://github.com/apache/tinkerpop}{TinkerPop} & 1698 & 95652 &  \multicolumn{1}{r}{5178} \\
        \href{https://github.com/apache/zeppelin}{Zeppelin} & 1209 & 89193 &  \multicolumn{1}{r}{1562} \\
        \bottomrule
    \end{tabular}
\end{table}

%% file: Results.tex
%!TEX root=bare_jrnl_compsoc.tex

\section{Results}\label{sec:results}

In this section we present the results of our study organized by research question, and highlight the major findings. However, prior to answering the research questions, we present a visualisation and descriptive statistics on the \TDd of individual commits for the selected projects. This can provide the context upon which we can interpret the results of the research questions.
%The rest of the discussion (i.e., comparison to related work, interpretation, and implications for researchers and practitioners) takes place in Section \ref{sec:discussion}.

\subsection{Descriptive Statistics}\label{subsec:descriptive}

\begin{figure}[!b]
    \centering
    \includegraphics[width=\linewidth]{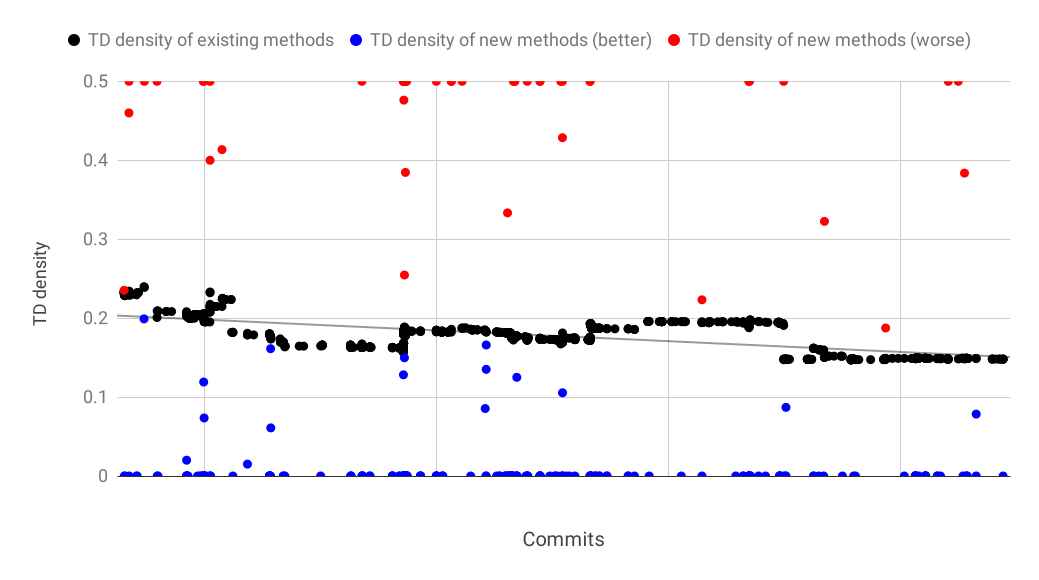}
    \caption{Contribution of new code on system's \TDd (Motivating Example)}
    \label{fig:rq1Chart}
\end{figure}

To obtain a first insight into the quality of new code as opposed to the quality of the system in which the new methods are added, we plot the evolution of the system's \TDd along with the \TDd of individual commits where new methods are added. 

Figure \ref{fig:rq1Chart} illustrates for one project (\project{Commons IO})\footnote{Project Commons IO is used as a motivating example and has not been included in our dataset since it is rather small and has a limited number of revisions} the evolution of the system's \TDd (black dots) and the corresponding trend-line, depicting a gradually increasing quality (black line declines over time). On the same plot, blue dots correspond to the revisions, in which the \TDd of new methods was lower than that of the system in that revision, while red dots indicate the cases where the \TDd of new methods was higher\footnote{For clarity, an upper bound on the displayed \TDd values is imposed, that is, data points with an extremely high \TDd are not accurately depicted but are simply placed on the upper bound (top of the figure).}. As it can be observed, for the vast majority of revisions (77\%), the \TDd of new methods is lower than the \TDd of the host system and in many cases the new code is entirely TD-free (see blue dots on the x-axis).

\input{tables/RQ1descriptive.tex}

As it is not possible to show similar plots for all projects, Table \ref{tab:rq1Descriptive} shows the percentage of revisions for which the \TDd of new code was lower than the \TDd of the system in the corresponding revision. The findings confirm the first impression that new code in the examined systems is generally of a higher quality than the existing baseline: in the majority of the commits, new methods have lower \TDd than the host system. Considering that many of these systems have a quality that increases over time, it would be reasonable to argue that the cleanness of new code has contributed to the declining trend of the system \TDd.  However, to claim that new code is a prominent factor that leads to the reduction of TD requires further analysis.

\begin{figure}[!t]
    \centering
    \includegraphics[width=\linewidth]{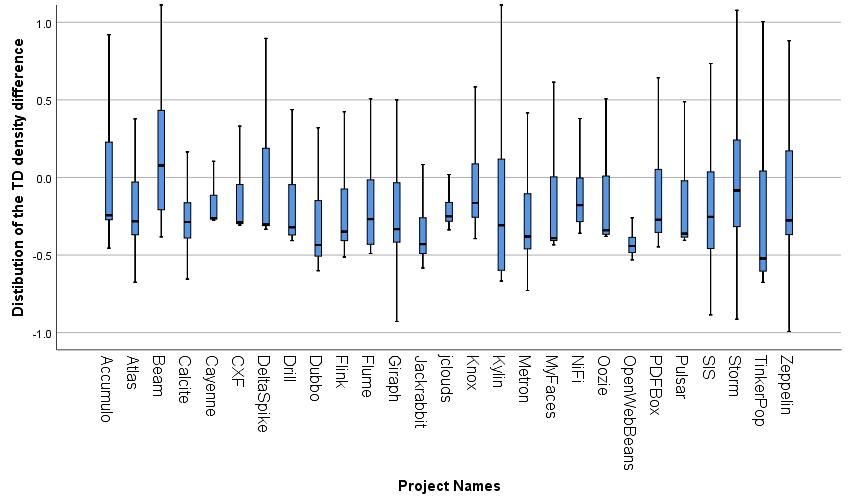}
    \caption{Distribution of the difference between the \TDd of new methods (introduced in new classes) and the \TDd of existing system, for all projects}
    \label{fig:rq1ADD}
\end{figure}

\begin{figure}[!t]
    \centering
    \includegraphics[width=\linewidth]{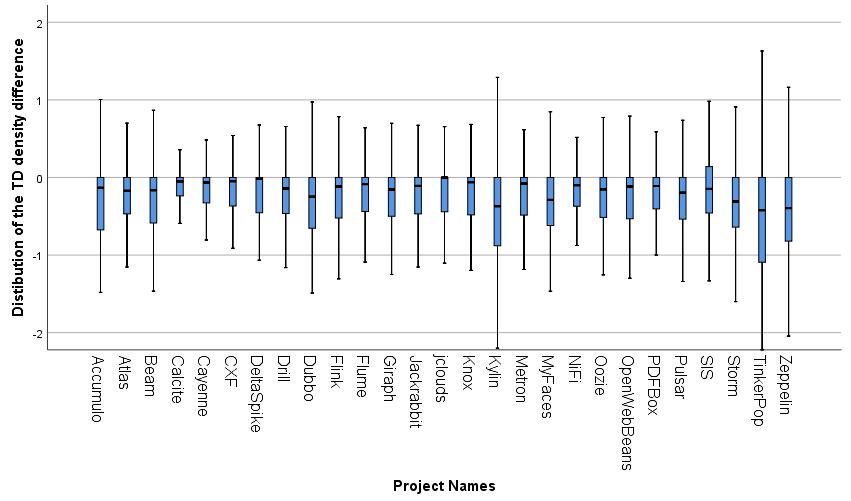}
    \caption{Distribution of the difference between the \TDd of new methods (introduced in existing classes) and the \TDd of existing system, for all projects}
    \label{fig:rq1MOD}
\end{figure}

Moving on to more detailed descriptive statistics, the boxplots in Fig. \ref{fig:rq1ADD} and Fig. \ref{fig:rq1MOD} illustrate the distribution of the difference between the \TDd of new methods and the \TDd density of the host code.
To allow for a fair comparison, we differentiate between the case when one or more new methods are introduced in an existing class and the case when a set of new methods are introduced in the form of a new class. In the former case the \TDd of the new methods should be contrasted against that of the class in which they are added, since the class resembles the \textit{neighborhood} of the new code, in terms of functionality and complexity. For the case of completely new classes, the comparison should be made against the entire system in which the new class is added, as the system is the \textit{neighborhood} of the introduced class. The boxplot of Fig. \ref{fig:rq1ADD} shows the distribution of the difference in \TDd between new methods added in \emph{new} classes at revision \textit{i} and the quality of the entire system in the previous revision (\textit{i-1}). The boxplot of Fig. \ref{fig:rq1MOD} shows the distribution of the difference in \TDd between new methods added in \emph{existing} classes at revision \textit{i} and the quality of the class in which they are added, in the previous (\textit{i-1}) revision. 

As it can be observed from the boxplots, the median difference ($\mu$) between the \TDd of new methods and that of the host code, is negative for all but one projects. For most of the projects, and especially for new methods introduced in existing classes, even the upper quartile is below zero. Thus, it becomes evident that in the transitions in which new code was added (in the form of entirely new methods either in new classes or in existing classes) the \TDd of the new code is significantly lower than that of the host code while in many cases it is very close to zero.

\input{tables/RQ2HeatmapTable.tex}

\subsection{Relation among the contribution of new / deleted / modified methods and change in system's \TDd (RQ$_{1}$)}\label{subsec:resRQ1}

% The goal of the first research question is to investigate how the different types of changes that can be performed at the method level (addition, deletion and modification), from one revision to the next, contribute to the changes in the project's \TDd. 
We remind that for studying this RQ we created two categorical variables: the first refers to the direction of change (decreasing/increasing\footnote{The cases where the system's 
\TDd remained stable are rare and are omitted for the sake of simplicity}) of the system's \TDd in each revision. The other refers to the contribution (decreasing, increasing, stable) of new/deleted/modified methods in each revision. The contribution itself is calculated according to equations (1)-(3). We have turned this contribution to a categorical variable depending on whether the contribution is positive, zero, or negative.

The results from the cross-tabulation of frequencies between these two variables are displayed in Table \ref{tab:rq2Heatmap} for all the analyzed projects. For each project (composite row) three individual heatmaps are presented, one for the contribution of new, deleted and modified code respectively. Each heatmap is comprised of six cells: the two rows correspond to the increase or decrease in the system's \TDd, whereas the three columns to the effect (decrease, increase and stable) of the corresponding change type (new, deleted or modified). The intensity of the color (within each 6-cell heatmap) indicates the frequency of occurrence for each combination between the two categorical variables, that is, the direction of change in the system's \TDd (decrease or increase), and the direction of change (decrease, increase and stable) for each type of contribution. It should be noted that while absolute numbers (number of transitions in which each combination has been observed) are shown on the heatmaps, the intensity of the colors reflects the corresponding percentage of cases (highest percentage corresponds to the most intense red). 
%The cell on which we focus in the current study is the top left for each type of contribution and project; this reveals the ability of new/deleted/modified code to contribute to the decrease of the overall \TDd. For example in the NEW column, a warm red color in that cell implies that new code that is of better quality most frequently coincides with a reduction of the system's \TDd.

Let us consider as an example, the first project in Table {\ref{tab:rq2Heatmap}} (project {\project{Accumulo}}). We focus on the contribution of new code (composite column New) and study separately the two rows, corresponding to transitions where the system's {\TDd} has decreased and increased, respectively:
\begin{itemize}
  \item \emph{System {\TDd} has decreased (top row)}: The warm (red) color in the upper left cell (labeled with 195) implies that among the cases where new code was added 
  %(competing with at least one other type of code change, i.e. method deletion and/or modification) 
  and the system {\TDd} has decreased, the highest frequency (195 out of the total 242 transitions) was observed for new methods that contributed to a decrease in the \TDd.
  \item \emph{System {\TDd} has increased (bottom row)}: The red color in the lower center cell (labeled with 113) implies that among the cases where new code was added and the system {\TDd} has increased, the highest frequency (113 out of the total 210 transitions) was observed for new methods that contributed to an increase in the \TDd.
\end{itemize}
In other words, in this project, the change in the system {\TDd} co-occured in most of the cases with a contribution of the same direction by new code.

\input{tables/RQ2.tex}

\begin{figure*}[!t]
    \centering
    \includegraphics[width=\linewidth]{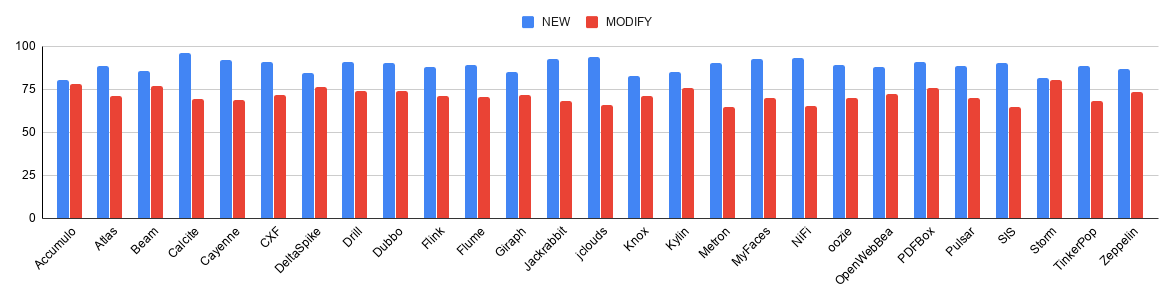}
    \caption{Percentage of revisions in which a decrease in the system \TDd co-occurred with a positive contribution in quality by new and modified methods}
    \label{fig:rq2barchart}
\end{figure*}

% In the majority of the projects, the consistency in the relationship (i.e. that system’s {\TDd} and new methods’{\TDd} change in the same direction) is observed for the decrease of {\TDd}, which is actually the desired effect for new code.
    
% Let us consider as an example, the fourth project in Table \ref{tab:rq2Heatmap} (project \project{Calcite}). The upper left cell in that row (labeled with 607)
% corresponds to the transitions for which a decrease (top row) in the system \TDd co-occurred with the introduction of new code that resulted in a decrease in the \TDd. The warm (red) color implies that among the cases where new code was added (competing with at least one other type of code change, i.e. method deletion and/or modification) and the system's 
% \TDd has decreased, the highest frequency (607 out of the total 634 transitions) was observed for new methods that contributed to a decrease in the \TDd. Looking at the second row for that project, we observe that among the cases where new code was added and the system's \TDd has increased, the highest frequency (209 out of the total 317 transitions) was observed for new methods that contributed to a decrease in the \TDd. However, as it can be seen from the color intensity, this frequency does not stand out as intensively as for the top row, i.e. in another 108 transitions where new code was added and the system's \TDd has increased, the new code has contributed to an increase of the \TDd.

Following this kind of interpreting Table \ref{tab:rq2Heatmap}, for the contribution of new methods in all projects, it can be observed that when their contribution leads to a reduction of \TDd, in most of the revisions the same direction of change is observed in the system's \TDd (warm red color in the top-left cell in each six-cell heatmap). In other words, in most of the cases, \textbf{when new code is \textit{cleaner}, the system's \TDd decreases}. However, an impact on the system's \TDd cannot be claimed  when new code contributes to an increase of the \TDd. It should be noted that this pattern is consistent among all projects. For deleted methods the most striking observation (most red cells) concerns the cases when the deleted methods contribute to an increase of the \TDd (for example, when high quality code is removed from the system). In those cases it seems that \textbf{the deletion of high quality code most frequently co-exists with an increase in the system's \TDd}.

An interesting and repeating pattern is present for modified methods. Intense red colors are observed in alternating rows: this implies that the direction of change in the system's \TDd coincides with the contribution of modified code. In other words, \textbf{if a method is modified and the \TDd of the method decreases, then, for the majority of the cases a decrease in the system's \TDd is observed; similarly for an increase in the \TDd}.

Finally, as expected, the lack of any contribution of new/deleted/modified methods (column --) does not appear to have any association with the overall change in the system's \TDd as depicted by the mostly blue cells. It should be emphasized, that these observations, should by no means be interpreted as indications of causality. Investigating whether each type of change (new/deleted/modified code) is responsible for the change in the overall \TDd would require a different experimental set up and is beyond the scope of this study. 

The chi-square test for independence has been used to discover if there is a relationship between the direction of change in the project's \TDd and the contribution of new/deleted/modified code. Table \ref{tab:rq2} shows for each project the Pearson chi-square value (top row) and whether the results are statistically significant or not depending on the p-value. The bottom row for each project shows the Phi value that tests the strength of the association \cite{pace2012beginning}.

As it can be observed, in almost all cases the results are statistically significant (p$<$0.01) implying that the null hypothesis can be rejected. In other words we can argue that there is a relationship between the direction of change in the systems \TDd and the contribution of new/deleted/modified code. In particular, \textbf{the strength of the association appears to be higher for the contribution of modified methods (for most projects), followed by the contribution of new code. The contribution of deleted methods seems to have a significantly lower association to the change in the system's \TDd}.

These results concern both directions of change in the system's \TDd and reveal that code modification can contribute positively and negatively to changes in the system quality. If we focus only on the cases where the \TDd decreased from one revision to the next, the potential of cleaner new methods becomes more evident: The barchart of Figure \ref{fig:rq2barchart} displays the percentage of transitions where a decrease of the system's \TDd was observed and new/modified methods also contributed to a decrease of \TDd (assuming that at least two types of changes were competing in the same transition). The cases where a positive contribution by new methods co-occurred with an improvement in the overall quality, are slightly more frequent.

\input{tables/RQ3.tex}

% OLD
% \begin{figure}[!h]
% \centering
% \begin{subfigure}[b]{0.95\linewidth}
%   \includegraphics[width=1\linewidth]{figures/bp1.png}
%   \caption{Distribution of the percentage of cleaner code commits for the two groups of projects, based on the existence of commit guidelines}
%   \label{fig:rq3BP1}
% \end{subfigure}
% \begin{subfigure}[b]{0.95\linewidth}
%   \includegraphics[width=1\linewidth]{figures/bp2.png}
%   \caption{Distribution of the percentage of cleaner code commits for the two groups of projects, based on the frequency of references to code quality in board meetings}
%   \label{fig:rq3BP2}
% \end{subfigure}
% \caption{Code quality practices and cleanness of new code}
% \end{figure}

\subsection{Relation between Code Quality Practices and New Code Cleanness
{(RQ$_{2}$)}}\label{subsec:resRQ2}
In Table {\ref{tab:rq3}}, we present the data extraction results for the studied projects, with respect to the employment of code quality practices at project management level.

\begin{figure}[!h]
    \centering
    \includegraphics[width=0.95\linewidth]{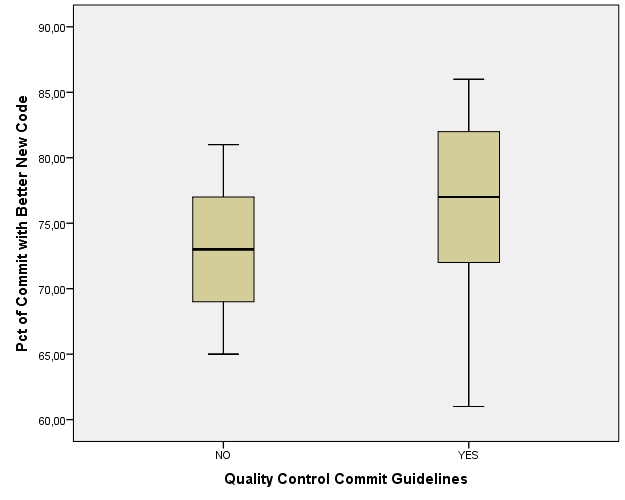}
    \caption{Distribution of the percentage of cleaner code commits for the two groups of projects, based on the existence of commit guidelines}
    \label{fig:rq3BP1}
\end{figure}

With respect to the existence of commit guidelines, we can observe that 16 projects provide guidelines related to TD rule violations, and 11 projects do not. By comparing the median values of the percentage of cleaner code commits (see Figure {\ref{fig:rq3BP1})} in the two groups, we can observe that projects that provide commit guidelines are having more commits (an increase of the median by 3.88\% has been observed) in which the new code is cleaner compared to existing code.  However, this difference is not statistically significant, based on the independent samples t-test (sig = 0.09).

\begin{figure}[!h]
    \centering
    \includegraphics[width=0.95\linewidth]{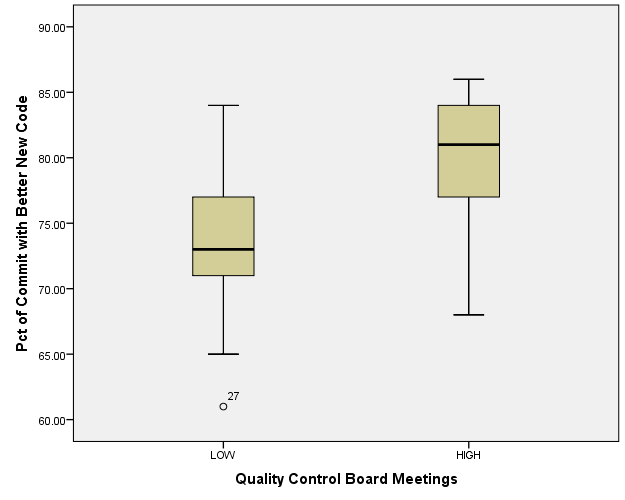}
    \caption{Distribution of the percentage of cleaner code commits for the two groups of projects, based on the frequency of references to code quality in board meetings}
    \label{fig:rq3BP2}
\end{figure}

On the other hand, \textbf{the 9 projects in which the project management team more regularly discusses code quality in the board meetings, are having a statistically significant higher percentage of commits in which the code is cleaner} (mean$_{diff}$ = 5\%, and sig = 0.05).  This difference is visualized in the boxplots of Figure {\ref{fig:rq3BP2}}, in which we can observe no overlap in the boxes (Q3-Q1) of the two groups.

%% file: tables/RQ1descriptive.tex
%!TEX root=bare_jrnl_compsoc.tex

\begin{table}[!t]
    \caption{Percentage of revisions in which new methods have lower \TDd, for all examined projects}
    \label{tab:rq1Descriptive}
    \begin{center}
    \begin{tabular}{l r | l r | l r}
    \toprule
    Project & \% & Project & \% & Project & \%\\
    \midrule
    \href{https://github.com/apache/accumulo}{Accumulo} & \rnd{0}{64.88549618}	&	\href{https://github.com/apache/flink}{Flink} & \rnd{0}{76.78812416}	&	\href{https://github.com/apache/nifi}{NiFi} & \rnd{0}{81.15942029}\\
    \href{https://github.com/apache/atlas}{Atlas} & \rnd{0}{72.55594817}	&	\href{https://github.com/apache/flume}{Flume} & \rnd{0}{72.87356322}	&	\href{https://github.com/apache/oozie}{oozie} & \rnd{0}{78.81619938}\\
    \href{https://github.com/apache/beam}{Beam} & \rnd{0}{70.62391681}	&	\href{https://github.com/apache/giraph}{Giraph} & \rnd{0}{74.04371585}	&	\href{https://github.com/apache/openwebbeans}{OpenWebBeans} & \rnd{0}{77.18023256}\\
    \href{https://github.com/apache/calcite}{Calcite} & \rnd{0}{86.06001936}	&	\href{https://github.com/apache/jackrabbit}{Jackrabbit} & \rnd{0}{80.85731063}	&	\href{https://github.com/apache/pdfbox}{PDFBox} & \rnd{0}{76.98795181}\\
    \href{https://github.com/apache/cayenne}{Cayenne} & \rnd{0}{82.3943662}	&	\href{https://github.com/apache/jclouds}{jclouds} & \rnd{0}{83.51324828}	&	\href{https://github.com/apache/pulsar}{Pulsar} & \rnd{0}{72.24938875}\\
    \href{https://github.com/apache/cxf}{CXF} & \rnd{0}{76.86292548}	&	\href{https://github.com/apache/knox}{Knox} & \rnd{0}{68.94736842}	&	\href{https://github.com/apache/sis}{SIS} & \rnd{0}{67.68292683}\\
    \href{https://github.com/apache/deltaspike}{DeltaSpike} & \rnd{0}{72.19917012}	&	\href{https://github.com/apache/kylin}{Kylin} & \rnd{0}{70.85597826}	&	\href{https://github.com/apache/storm}{Storm} & \rnd{0}{61.40350877}\\
    \href{https://github.com/apache/drill}{Drill} & \rnd{0}{84.39490446}	&	\href{https://github.com/apache/metron}{Metron} & \rnd{0}{84.35013263}	&	\href{https://github.com/apache/tinkerpop}{TinkerPop} & \rnd{0}{74.81927711}\\
    \href{https://github.com/apache/incubator-dubbo}{Dubbo} & \rnd{0}{80.86642599}	&	\href{https://github.com/apache/myfaces}{MyFaces} & \rnd{0}{71.16564417}	&	\href{https://github.com/apache/zeppelin}{Zeppelin} & \rnd{0}{71.50063052}\\
    \bottomrule
    \end{tabular}
    \end{center}
\end{table}

%% file: tables/RQ2HeatmapTable.tex
%!TEX root=bare_jrnl_compsoc.tex

\newcolumntype{P}[1]{>{\centering\arraybackslash}p{#1}}

\begin{table*}[!ht]
    \caption{Relation between contribution of new/deleted/modified methods and change in system's \TDd (RQ$_{1}$)}
    \label{tab:rq2Heatmap}
    \begin{center}
% The standard font size switches are: \tiny, \scriptsize, \footnotesize, \small, \normalsize, \large, \Large, \LARGE, \huge, and \Huge
% \large
\begin{tabular}{llccc|ccc|ccc}
\toprule

\multicolumn{2}{c}{\multirow{3}{*}{TD Density Change per Project}} & \multicolumn{9}{c}{Contribution to TD Density per new/delete/modified methods} \\ \cline{3-11}
\multicolumn{2}{c}{} & \multicolumn{3}{c}{New} & \multicolumn{3}{c}{Deleted} & \multicolumn{3}{c}{Modified} \\ \cline{3-11}
\multicolumn{2}{p{.05\textwidth}}{} & \multicolumn{1}{P{.06\textwidth}}{$\downarrow$} & \multicolumn{1}{P{.06\textwidth}}{$\uparrow$} & \multicolumn{1}{P{.06\textwidth}}{-} & \multicolumn{1}{P{.06\textwidth}}{$\downarrow$} & \multicolumn{1}{P{.06\textwidth}}{$\uparrow$} & \multicolumn{1}{P{.06\textwidth}}{-} & \multicolumn{1}{P{.06\textwidth}}{$\downarrow$} & \multicolumn{1}{P{.06\textwidth}}{$\uparrow$} &\multicolumn{1}{P{.06\textwidth}}{-} \\ \hline

% {\cellcolor[HTML]
\multirow{2}{*}{\href{https://github.com/apache/accumulo}{Accumulo}} & $\downarrow$ & {\cellcolor[HTML]{f8696b} 195} & {\cellcolor[HTML]{b9cde6} 47} & {\cellcolor[HTML]{598ac5} 0}  & {\cellcolor[HTML]{fbcdd0} 51} & {\cellcolor[HTML]{fccbce} 52} & {\cellcolor[HTML]{5f8dc8} 1} & {\cellcolor[HTML]{f8696b} 205} & {\cellcolor[HTML]{cddcef} 39} & {\cellcolor[HTML]{83a8d5} 18}\\

& $\uparrow$ & {\cellcolor[HTML]{fbd2d6} 97} & {\cellcolor[HTML]{fabcbd} 113} & {\cellcolor[HTML]{598ac5} 0}  & {\cellcolor[HTML]{b3c8e5} 15} & {\cellcolor[HTML]{f8696b} 66} & {\cellcolor[HTML]{598ac5} 0} & {\cellcolor[HTML]{fcf0f2} 55} & {\cellcolor[HTML]{f97779} 163} & {\cellcolor[HTML]{598ac5} 5}\\ \hline

\multirow{2}{*}{\href{https://github.com/apache/atlas}{Atlas}} & $\downarrow$ & {\cellcolor[HTML]{f8696b} 446} & {\cellcolor[HTML]{96b3db} 56} & {\cellcolor[HTML]{598ac5} 1}  & {\cellcolor[HTML]{fbeff1} 73} & {\cellcolor[HTML]{f98284} 156} & {\cellcolor[HTML]{598ac5} 2} & {\cellcolor[HTML]{f97d7f} 373} & {\cellcolor[HTML]{fcf7fb} 126} & {\cellcolor[HTML]{6f98cc} 25}\\

& $\uparrow$ & {\cellcolor[HTML]{fbced1} 141} & {\cellcolor[HTML]{fbc4c7} 153} & {\cellcolor[HTML]{5d8bc6} 2}  & {\cellcolor[HTML]{e1e9f4} 31} & {\cellcolor[HTML]{f8696b} 102} & {\cellcolor[HTML]{5e8cc7} 2} & {\cellcolor[HTML]{eaeef9} 62} & {\cellcolor[HTML]{f8696b} 249} & {\cellcolor[HTML]{598ac5} 7}\\ \hline

\multirow{2}{*}{\href{https://github.com/apache/beam}{Beam}} & $\downarrow$ & {\cellcolor[HTML]{f8696b} 550} & {\cellcolor[HTML]{a0badf} 86} & {\cellcolor[HTML]{598ac5} 8} & {\cellcolor[HTML]{fde4e7} 137} & {\cellcolor[HTML]{fa9a9c} 235} & {\cellcolor[HTML]{598ac5} 11} & {\cellcolor[HTML]{f8696b} 542} & {\cellcolor[HTML]{ecf0f9} 131} & {\cellcolor[HTML]{608ec9} 33}\\

& $\uparrow$ & {\cellcolor[HTML]{fbd2d6} 172} & {\cellcolor[HTML]{fbc0c4} 198} & {\cellcolor[HTML]{5f8dc8} 8} & {\cellcolor[HTML]{c3d3ea} 44} & {\cellcolor[HTML]{f8696b} 184} & {\cellcolor[HTML]{5e8cc7} 8} & {\cellcolor[HTML]{fbf9fc} 90} & {\cellcolor[HTML]{f87072} 311} & {\cellcolor[HTML]{598ac5} 17}\\ \hline

\multirow{2}{*}{\href{https://github.com/apache/calcite}{Calcite}} & $\downarrow$ & {\cellcolor[HTML]{f8696b} 607} & {\cellcolor[HTML]{7da3d2} 27} & {\cellcolor[HTML]{598ac5} 0} & {\cellcolor[HTML]{fdf8fc} 44} & {\cellcolor[HTML]{f96d6e} 157} & {\cellcolor[HTML]{598ac5} 3} & {\cellcolor[HTML]{f98b8c} 446} & {\cellcolor[HTML]{fcf1f5} 165} & {\cellcolor[HTML]{7ea2d2} 32}\\

& $\uparrow$ & {\cellcolor[HTML]{f9a3a4} 209} & {\cellcolor[HTML]{fde0e4} 108} & {\cellcolor[HTML]{598ac5} 0} & {\cellcolor[HTML]{ebeff8} 23} & {\cellcolor[HTML]{f8696b} 102} & {\cellcolor[HTML]{6f98cc} 5} & {\cellcolor[HTML]{d5e1f1} 53} & {\cellcolor[HTML]{f8696b} 277} & {\cellcolor[HTML]{598ac5} 1}\\ \hline

\multirow{2}{*}{\href{https://github.com/apache/cayenne}{Cayenne}} & $\downarrow$ & {\cellcolor[HTML]{f8696b} 528} & {\cellcolor[HTML]{8fafd8} 46} & {\cellcolor[HTML]{598ac5} 2} & {\cellcolor[HTML]{fce8ea} 88} & {\cellcolor[HTML]{f98f91} 202} & {\cellcolor[HTML]{598ac5} 2} & {\cellcolor[HTML]{f98689} 421} & {\cellcolor[HTML]{fbf5f7} 135} & {\cellcolor[HTML]{98b6dc} 58}\\

& $\uparrow$ & {\cellcolor[HTML]{fbacaf} 173} & {\cellcolor[HTML]{fbdbde} 111} & {\cellcolor[HTML]{598ac5} 1} & {\cellcolor[HTML]{b4c9e6} 22} & {\cellcolor[HTML]{f8696b} 155} & {\cellcolor[HTML]{6592c9} 4} & {\cellcolor[HTML]{dbe4f3} 52} & {\cellcolor[HTML]{f8696b} 269} & {\cellcolor[HTML]{598ac5} 12}\\ \hline

\multirow{2}{*}{\href{https://github.com/apache/cxf}{CXF}} & $\downarrow$ & {\cellcolor[HTML]{f8696b} 119} & {\cellcolor[HTML]{8bacd7} 109} & {\cellcolor[HTML]{598ac5} 7} & {\cellcolor[HTML]{fdf4f7} 90} & {\cellcolor[HTML]{f97b7e} 285} & {\cellcolor[HTML]{598ac5} 7} & {\cellcolor[HTML]{f9797a} 917} & {\cellcolor[HTML]{fcfcfe} 262} & {\cellcolor[HTML]{8bacd7} 97}\\

& $\uparrow$ & {\cellcolor[HTML]{fbc3c6} 344} & {\cellcolor[HTML]{fbced1} 310} & {\cellcolor[HTML]{598ac5} 3} & {\cellcolor[HTML]{dbe4f3} 35} & {\cellcolor[HTML]{f8696b} 178} & {\cellcolor[HTML]{598ac5} 4} & {\cellcolor[HTML]{faf9fe} 142} & {\cellcolor[HTML]{f8696b} 548} & {\cellcolor[HTML]{598ac5} 15}\\ \hline

\multirow{2}{*}{\href{https://github.com/apache/deltaspike}{DeltaSpike}} & $\downarrow$ & {\cellcolor[HTML]{f8696b} 102} & {\cellcolor[HTML]{a5bddf} 18} & {\cellcolor[HTML]{598ac5} 1} & {\cellcolor[HTML]{fbcdd0} 21} & {\cellcolor[HTML]{fbb9bb} 24} & {\cellcolor[HTML]{598ac5} 2} & {\cellcolor[HTML]{f8696b} 95} & {\cellcolor[HTML]{acc3e2} 15} & {\cellcolor[HTML]{acc3e2} 15}\\

& $\uparrow$ & {\cellcolor[HTML]{fbc4c7} 39} & {\cellcolor[HTML]{fad0d2} 36} & {\cellcolor[HTML]{598ac5} 1} & {\cellcolor[HTML]{9ebadf} 5} & {\cellcolor[HTML]{f8696b} 23} & {\cellcolor[HTML]{799fd0} 3} & {\cellcolor[HTML]{fde4e7} 25} & {\cellcolor[HTML]{f98183} 55} & {\cellcolor[HTML]{598ac5} 2}\\ \hline

\multirow{2}{*}{\href{https://github.com/apache/drill}{Drill}} & $\downarrow$ & {\cellcolor[HTML]{f8696b} 521} & {\cellcolor[HTML]{a9c2e1} 51} & {\cellcolor[HTML]{598ac5} 1} & {\cellcolor[HTML]{fdeaec} 77} & {\cellcolor[HTML]{f98c8f} 181} & {\cellcolor[HTML]{598ac5} 6} & {\cellcolor[HTML]{f98b8c} 427} & {\cellcolor[HTML]{fcf0f2} 118} & {\cellcolor[HTML]{98b6dc} 35}\\

& $\uparrow$ & {\cellcolor[HTML]{f88e90} 212} & {\cellcolor[HTML]{fdeaec} 78} & {\cellcolor[HTML]{5a8bc6} 1} & {\cellcolor[HTML]{b9cde6} 17} & {\cellcolor[HTML]{f8696b} 104} & {\cellcolor[HTML]{6190c8} 4} & {\cellcolor[HTML]{adc4e3} 23} & {\cellcolor[HTML]{f8696b} 274} & {\cellcolor[HTML]{598ac5} 3}\\ \hline

\multirow{2}{*}{\href{https://github.com/apache/incubator-dubbo}{Dubbo}} & $\downarrow$ & {\cellcolor[HTML]{f8696b} 144} & {\cellcolor[HTML]{a1bbde} 15} & {\cellcolor[HTML]{598ac5} 1} & {\cellcolor[HTML]{fbe5e7} 28} & {\cellcolor[HTML]{f9888a} 62} & {\cellcolor[HTML]{598ac5} 2} & {\cellcolor[HTML]{f9888a} 123} & {\cellcolor[HTML]{fdf4f7} 32} & {\cellcolor[HTML]{a6c0e1} 12}\\

& $\uparrow$ & {\cellcolor[HTML]{faa0a2} 54} & {\cellcolor[HTML]{fde4e7} 27} & {\cellcolor[HTML]{719ace} 3} & {\cellcolor[HTML]{a8c1e0} 5} & {\cellcolor[HTML]{f8696b} 35} & {\cellcolor[HTML]{94b4db} 4} & {\cellcolor[HTML]{d0ddee} 10} & {\cellcolor[HTML]{f8696b} 80} & {\cellcolor[HTML]{598ac5} 0}\\ \hline

\multirow{2}{*}{\href{https://github.com/apache/flink}{Flink}} & $\downarrow$ & {\cellcolor[HTML]{f8696b} 1560} & {\cellcolor[HTML]{9bb7dc} 210} & {\cellcolor[HTML]{598ac5} 9} & {\cellcolor[HTML]{fcf0f2} 307} & {\cellcolor[HTML]{f98183} 673} & {\cellcolor[HTML]{598ac5} 20} & {\cellcolor[HTML]{f9797a} 1314} & {\cellcolor[HTML]{fbf9fc} 465} & {\cellcolor[HTML]{6a95ca} 76}\\

& $\uparrow$ & {\cellcolor[HTML]{fbbdc0} 477} & {\cellcolor[HTML]{fbd3d4} 402} & {\cellcolor[HTML]{598ac5} 5} & {\cellcolor[HTML]{e1e9f4} 126} & {\cellcolor[HTML]{f8696b} 416} & {\cellcolor[HTML]{5d8bc6} 13} & {\cellcolor[HTML]{f1f4fb} 222} & {\cellcolor[HTML]{f8696b} 756} & {\cellcolor[HTML]{598ac5} 18}\\ \hline

\multirow{2}{*}{\href{https://github.com/apache/flume}{Flume}} & $\downarrow$ & {\cellcolor[HTML]{f8696b} 224} & {\cellcolor[HTML]{96b3db} 25} & {\cellcolor[HTML]{5f8dc8} 2} & {\cellcolor[HTML]{f7f8fd} 21} & {\cellcolor[HTML]{f8696b} 67} & {\cellcolor[HTML]{6190c8} 1} & {\cellcolor[HTML]{fa797d} 178} & {\cellcolor[HTML]{fcfcfe} 59} & {\cellcolor[HTML]{7ea2d2} 16}\\

& $\uparrow$ & {\cellcolor[HTML]{fcd5d8} 53} & {\cellcolor[HTML]{fbb6b9} 69} & {\cellcolor[HTML]{598ac5} 0} & {\cellcolor[HTML]{fcfafd} 11} & {\cellcolor[HTML]{f96a6c} 33} & {\cellcolor[HTML]{598ac5} 0} & {\cellcolor[HTML]{f8f9fd} 29} & {\cellcolor[HTML]{f8696b} 99} & {\cellcolor[HTML]{598ac5} 2}\\ \hline

\multirow{2}{*}{\href{https://github.com/apache/giraph}{Giraph}} & $\downarrow$ & {\cellcolor[HTML]{f8696b} 172} & {\cellcolor[HTML]{a7bfe1} 29} & {\cellcolor[HTML]{598ac5} 1} & {\cellcolor[HTML]{fce8ea} 31} & {\cellcolor[HTML]{fa9396} 67} & {\cellcolor[HTML]{598ac5} 2} & {\cellcolor[HTML]{fa8385} 143} & {\cellcolor[HTML]{fdf4f7} 48} & {\cellcolor[HTML]{769ed1} 9}\\

& $\uparrow$ & {\cellcolor[HTML]{fabec0} 56} & {\cellcolor[HTML]{fbd4d7} 47} & {\cellcolor[HTML]{6190c8} 2} & {\cellcolor[HTML]{b2c7e4} 5} & {\cellcolor[HTML]{f8696b} 33} & {\cellcolor[HTML]{5f8dc8} 1} & {\cellcolor[HTML]{dfe9f5} 19} & {\cellcolor[HTML]{f8696b} 91} & {\cellcolor[HTML]{598ac5} 1}\\ \hline

\multirow{2}{*}{\href{https://github.com/apache/jackrabbit}{Jackrabbit}} & $\downarrow$ & {\cellcolor[HTML]{f8696b} 915} & {\cellcolor[HTML]{88aad7} 74} & {\cellcolor[HTML]{598ac5} 1} & {\cellcolor[HTML]{fcf1f5} 111} & {\cellcolor[HTML]{f97d7f} 298} & {\cellcolor[HTML]{598ac5} 7} & {\cellcolor[HTML]{f98284} 711} & {\cellcolor[HTML]{fcf7fb} 245} & {\cellcolor[HTML]{87a9d6} 89}\\

& $\uparrow$ & {\cellcolor[HTML]{fbbdc0} 212} & {\cellcolor[HTML]{fcd5d8} 170} & {\cellcolor[HTML]{6390c9} 6} & {\cellcolor[HTML]{dde6f5} 35} & {\cellcolor[HTML]{f8696b} 149} & {\cellcolor[HTML]{5e8cc7} 4} & {\cellcolor[HTML]{e7eef8} 86} & {\cellcolor[HTML]{f8696b} 345} & {\cellcolor[HTML]{598ac5} 16}\\ \hline

\multirow{2}{*}{\href{https://github.com/apache/jclouds}{jclouds}} & $\downarrow$ & {\cellcolor[HTML]{f8696b} 1109} & {\cellcolor[HTML]{7ca2d1} 66} & {\cellcolor[HTML]{598ac5} 7} & {\cellcolor[HTML]{fdeef1} 136} & {\cellcolor[HTML]{f98689} 433} & {\cellcolor[HTML]{5f8dc8} 14} & {\cellcolor[HTML]{f98689} 821} & {\cellcolor[HTML]{fcf6f8} 294} & {\cellcolor[HTML]{95b2da} 131}\\

& $\uparrow$ & {\cellcolor[HTML]{fcb7ba} 272} & {\cellcolor[HTML]{fad6d8} 200} & {\cellcolor[HTML]{5f8dc8} 6} & {\cellcolor[HTML]{aec5e4} 29} & {\cellcolor[HTML]{f8696b} 268} & {\cellcolor[HTML]{598ac5} 6} & {\cellcolor[HTML]{e4eaf6} 105} & {\cellcolor[HTML]{f8696b} 429} & {\cellcolor[HTML]{598ac5} 24}\\ \hline

\multirow{2}{*}{\href{https://github.com/apache/knox}{Knox}} & $\downarrow$ & {\cellcolor[HTML]{f8696b} 271} & {\cellcolor[HTML]{8aabd6} 35} & {\cellcolor[HTML]{6a95ca} 21} & {\cellcolor[HTML]{fdeaec} 44} & {\cellcolor[HTML]{f97374} 68} & {\cellcolor[HTML]{598ac5} 2} & {\cellcolor[HTML]{f8696b} 241} & {\cellcolor[HTML]{dfe9f5} 74} & {\cellcolor[HTML]{799fd0} 24}\\

& $\uparrow$ & {\cellcolor[HTML]{fbd4d7} 68} & {\cellcolor[HTML]{fbb4b8} 88} & {\cellcolor[HTML]{598ac5} 7} & {\cellcolor[HTML]{ecf0f9} 23} & {\cellcolor[HTML]{f8696b} 44} & {\cellcolor[HTML]{739bce} 5} & {\cellcolor[HTML]{fbeff1} 55} & {\cellcolor[HTML]{f87678} 124} & {\cellcolor[HTML]{598ac5} 5}\\ \hline

\multirow{2}{*}{\href{https://github.com/apache/kylin}{Kylin}} & $\downarrow$ & {\cellcolor[HTML]{f8696b} 714} & {\cellcolor[HTML]{a5bfe0} 127} & {\cellcolor[HTML]{598ac5} 1} & {\cellcolor[HTML]{fde4e7} 177} & {\cellcolor[HTML]{fa9698} 295} & {\cellcolor[HTML]{598ac5} 4} & {\cellcolor[HTML]{f96e71} 688} & {\cellcolor[HTML]{f9fafe} 174} & {\cellcolor[HTML]{719ace} 47}\\

& $\uparrow$ & {\cellcolor[HTML]{faccce} 255} & {\cellcolor[HTML]{facccf} 254} & {\cellcolor[HTML]{5e8cc7} 4} & {\cellcolor[HTML]{cfdcef} 60} & {\cellcolor[HTML]{f8696b} 211} & {\cellcolor[HTML]{6390c9} 7} & {\cellcolor[HTML]{fcfcfe} 116} & {\cellcolor[HTML]{f8696b} 460} & {\cellcolor[HTML]{598ac5} 16}\\ \hline

\multirow{2}{*}{\href{https://github.com/apache/metron}{Metron}} & $\downarrow$ & {\cellcolor[HTML]{f8696b} 206} & {\cellcolor[HTML]{a6bee0} 21} & {\cellcolor[HTML]{5e8cc7} 1} & {\cellcolor[HTML]{eff4fa} 30} & {\cellcolor[HTML]{f8696b} 76} & {\cellcolor[HTML]{5f8dc8} 1} & {\cellcolor[HTML]{f99da0} 147} & {\cellcolor[HTML]{fce8ea} 72} & {\cellcolor[HTML]{769ed1} 9}\\

& $\uparrow$ & {\cellcolor[HTML]{fa9396} 71} & {\cellcolor[HTML]{fce8ea} 30} & {\cellcolor[HTML]{598ac5} 0} & {\cellcolor[HTML]{fdf4f7} 15} & {\cellcolor[HTML]{f97779} 31} & {\cellcolor[HTML]{598ac5} 0} & {\cellcolor[HTML]{b5cae7} 13} & {\cellcolor[HTML]{f8696b} 92} & {\cellcolor[HTML]{598ac5} 0}\\ \hline

\multirow{2}{*}{\href{https://github.com/apache/myfaces}{MyFaces}} & $\downarrow$ & {\cellcolor[HTML]{f8696b} 205} & {\cellcolor[HTML]{8daed9} 17} & {\cellcolor[HTML]{598ac5} 0} & {\cellcolor[HTML]{fcdfe1} 37} & {\cellcolor[HTML]{f99fa1} 57} & {\cellcolor[HTML]{598ac5} 3} & {\cellcolor[HTML]{f97b7e} 176} & {\cellcolor[HTML]{fcfafd} 56} & {\cellcolor[HTML]{87a9d6} 21}\\

& $\uparrow$ & {\cellcolor[HTML]{fcd9dd} 65} & {\cellcolor[HTML]{faafb3} 96} & {\cellcolor[HTML]{598ac5} 0} & {\cellcolor[HTML]{bfd1e9} 11} & {\cellcolor[HTML]{f8696b} 44} & {\cellcolor[HTML]{6893ca} 3} & {\cellcolor[HTML]{f5f6fb} 37} & {\cellcolor[HTML]{f8696b} 138} & {\cellcolor[HTML]{598ac5} 6}\\ \hline

\multirow{2}{*}{\href{https://github.com/apache/nifi}{NiFi}} & $\downarrow$ & {\cellcolor[HTML]{f8696b} 473} & {\cellcolor[HTML]{83a8d5} 32} & {\cellcolor[HTML]{5d8bc6} 2} & {\cellcolor[HTML]{fcedf0} 50} & {\cellcolor[HTML]{f98689} 113} & {\cellcolor[HTML]{598ac5} 3} & {\cellcolor[HTML]{f98b8e} 337} & {\cellcolor[HTML]{fbf5f7} 144} & {\cellcolor[HTML]{81a6d3} 36}\\

& $\uparrow$ & {\cellcolor[HTML]{fab4b6} 136} & {\cellcolor[HTML]{fbd7d9} 97} & {\cellcolor[HTML]{598ac5} 0} & {\cellcolor[HTML]{d3def0} 16} & {\cellcolor[HTML]{f8696b} 67} & {\cellcolor[HTML]{5e8cc7} 2} & {\cellcolor[HTML]{ebeff8} 55} & {\cellcolor[HTML]{f8696b} 188} & {\cellcolor[HTML]{598ac5} 2}\\ \hline

\multirow{2}{*}{\href{https://github.com/apache/oozie}{oozie}} & $\downarrow$ & {\cellcolor[HTML]{f8696b} 131} & {\cellcolor[HTML]{9fbcde} 22} & {\cellcolor[HTML]{5e8cc7} 1}  & {\cellcolor[HTML]{fce8ea} 30} & {\cellcolor[HTML]{fa9396} 58} & {\cellcolor[HTML]{598ac5} 0} & {\cellcolor[HTML]{f98183} 148} & {\cellcolor[HTML]{fcf7fb} 51} & {\cellcolor[HTML]{81a6d3} 13}\\

& $\uparrow$ & {\cellcolor[HTML]{faaaad} 60} & {\cellcolor[HTML]{fbdbde} 39} & {\cellcolor[HTML]{598ac5} 0}  & {\cellcolor[HTML]{cbdaed} 5} & {\cellcolor[HTML]{f8696b} 22} & {\cellcolor[HTML]{598ac5} 0} & {\cellcolor[HTML]{ecf0f9} 21} & {\cellcolor[HTML]{f8696b} 83} & {\cellcolor[HTML]{598ac5} 1}\\ \hline

\multirow{2}{*}{\href{https://github.com/apache/openwebbeans}{OpenWebBeans}} & $\downarrow$ & {\cellcolor[HTML]{f8696b} 330} & {\cellcolor[HTML]{8fafd8} 38} & {\cellcolor[HTML]{598ac5} 8} & {\cellcolor[HTML]{fcdcdf} 71} & {\cellcolor[HTML]{faa1a3} 107} & {\cellcolor[HTML]{598ac5} 3} & {\cellcolor[HTML]{f97b7e} 308} & {\cellcolor[HTML]{fcfafd} 88} & {\cellcolor[HTML]{88aad7} 31}\\

& $\uparrow$ & {\cellcolor[HTML]{fbb9bb} 95} & {\cellcolor[HTML]{fad6d8} 74} & {\cellcolor[HTML]{5f8dc8} 5} & {\cellcolor[HTML]{b9cde6} 17} & {\cellcolor[HTML]{f8696b} 76} & {\cellcolor[HTML]{6e97cd} 5} & {\cellcolor[HTML]{f1f4fb} 39} & {\cellcolor[HTML]{f8696b} 167} & {\cellcolor[HTML]{598ac5} 5}\\ \hline

\multirow{2}{*}{\href{https://github.com/apache/pdfbox}{PDFBox}} & $\downarrow$ & {\cellcolor[HTML]{f8696b} 385} & {\cellcolor[HTML]{8cadd8} 38} & {\cellcolor[HTML]{5c8ac6} 1} & {\cellcolor[HTML]{fddade} 60} & {\cellcolor[HTML]{fbacaf} 85} & {\cellcolor[HTML]{598bc6} 1} & {\cellcolor[HTML]{f8696b} 345} & {\cellcolor[HTML]{c9d8ed} 66} & {\cellcolor[HTML]{97b5db} 46}\\

& $\uparrow$ & {\cellcolor[HTML]{facccf} 107} & {\cellcolor[HTML]{fbc9cc} 110} & {\cellcolor[HTML]{598ac5} 0} & {\cellcolor[HTML]{b3c8e5} 13} & {\cellcolor[HTML]{f8696b} 68} & {\cellcolor[HTML]{5e8cc7} 1} & {\cellcolor[HTML]{fcf1f5} 56} & {\cellcolor[HTML]{f97374} 172} & {\cellcolor[HTML]{598ac5} 11}\\ \hline

\multirow{2}{*}{\href{https://github.com/apache/pulsar}{Pulsar}} & $\downarrow$ & {\cellcolor[HTML]{f8696b} 427} & {\cellcolor[HTML]{9ab8dc} 54} & {\cellcolor[HTML]{598ac5} 1} & {\cellcolor[HTML]{fcf1f5} 44} & {\cellcolor[HTML]{f98689} 100} & {\cellcolor[HTML]{6a95ca} 4} & {\cellcolor[HTML]{f9797a} 350} & {\cellcolor[HTML]{fcfcfe} 117} & {\cellcolor[HTML]{799fd0} 33}\\

& $\uparrow$ & {\cellcolor[HTML]{fbd4d5} 117} & {\cellcolor[HTML]{fbbfc1} 140} & {\cellcolor[HTML]{6190c8} 4} & {\cellcolor[HTML]{e3e9f5} 24} & {\cellcolor[HTML]{f8696b} 78} & {\cellcolor[HTML]{598ac5} 0} & {\cellcolor[HTML]{f7f8fd} 63} & {\cellcolor[HTML]{f8696b} 214} & {\cellcolor[HTML]{598ac5} 8}\\ \hline

\multirow{2}{*}{\href{https://github.com/apache/sis}{SIS}} & $\downarrow$ & {\cellcolor[HTML]{f8696b} 205} & {\cellcolor[HTML]{90b0d9} 22} & {\cellcolor[HTML]{598ac5} 1} & {\cellcolor[HTML]{fdeef1} 52} & {\cellcolor[HTML]{f98284} 98} & {\cellcolor[HTML]{598ac5} 1} & {\cellcolor[HTML]{f98c8f} 152} & {\cellcolor[HTML]{fcf0f2} 74} & {\cellcolor[HTML]{6d96cc} 10}\\

& $\uparrow$ & {\cellcolor[HTML]{fbd1d3} 81} & {\cellcolor[HTML]{fbc0c4} 94} & {\cellcolor[HTML]{598ac5} 1} & {\cellcolor[HTML]{e5ebf7} 31} & {\cellcolor[HTML]{f8696b} 84} & {\cellcolor[HTML]{608ec9} 2} & {\cellcolor[HTML]{e2eaf5} 44} & {\cellcolor[HTML]{f8696b} 144} & {\cellcolor[HTML]{598ac5} 2}\\ \hline

\multirow{2}{*}{\href{https://github.com/apache/storm}{Storm}} & $\downarrow$ & {\cellcolor[HTML]{f8696b} 133} & {\cellcolor[HTML]{b7cbe6} 28} & {\cellcolor[HTML]{5e8cc7} 2} & {\cellcolor[HTML]{fcf1f5} 37} & {\cellcolor[HTML]{faa0a2} 47} & {\cellcolor[HTML]{6d96cc} 4} & {\cellcolor[HTML]{f8696b} 136} & {\cellcolor[HTML]{cfdcef} 25} & {\cellcolor[HTML]{7ca0d0} 8}\\

& $\uparrow$ & {\cellcolor[HTML]{fbdbde} 61} & {\cellcolor[HTML]{faabae} 86} & {\cellcolor[HTML]{598ac5} 1} & {\cellcolor[HTML]{f6f7fc} 30} & {\cellcolor[HTML]{f8696b} 47} & {\cellcolor[HTML]{598ac5} 0} & {\cellcolor[HTML]{fbeff1} 40} & {\cellcolor[HTML]{f87a7d} 116} & {\cellcolor[HTML]{598ac5} 1}\\ \hline

\multirow{2}{*}{\href{https://github.com/apache/tinkerpop}{TinkerPop}} & $\downarrow$ & {\cellcolor[HTML]{f8696b} 1170} & {\cellcolor[HTML]{97b5db} 151} & {\cellcolor[HTML]{598ac5} 2} & {\cellcolor[HTML]{fcedf0} 200} & {\cellcolor[HTML]{f98183} 477} & {\cellcolor[HTML]{598ac5} 23} & {\cellcolor[HTML]{f98284} 949} & {\cellcolor[HTML]{fcfafd} 299} & {\cellcolor[HTML]{9fbcde} 145}\\

& $\uparrow$ & {\cellcolor[HTML]{fac8c9} 414} & {\cellcolor[HTML]{fbced1} 393} & {\cellcolor[HTML]{5d8bc6} 5} & {\cellcolor[HTML]{cfdcef} 80} & {\cellcolor[HTML]{f8696b} 350} & {\cellcolor[HTML]{6d96cc} 25} & {\cellcolor[HTML]{f2f5fc} 174} & {\cellcolor[HTML]{f8696b} 696} & {\cellcolor[HTML]{598ac5} 25}\\ \hline

\multirow{2}{*}{\href{https://github.com/apache/zeppelin}{Zeppelin}} & $\downarrow$ & {\cellcolor[HTML]{f86a69} 368} & {\cellcolor[HTML]{a2bcdf} 57} & {\cellcolor[HTML]{598ac5} 0} & {\cellcolor[HTML]{fcdcdf} 68} & {\cellcolor[HTML]{faa5a8} 91} & {\cellcolor[HTML]{598ac5} 3} & {\cellcolor[HTML]{f97374} 332} & {\cellcolor[HTML]{fcfcfe} 93} & {\cellcolor[HTML]{759dd0} 27}\\

& $\uparrow$ & {\cellcolor[HTML]{fcc1c5} 155} & {\cellcolor[HTML]{fbd1d3} 137} & {\cellcolor[HTML]{5a8bc6} 1} & {\cellcolor[HTML]{d0ddee} 22} & {\cellcolor[HTML]{f8696b} 63} & {\cellcolor[HTML]{6190c8} 3} & {\cellcolor[HTML]{f9f8fd} 61} & {\cellcolor[HTML]{f8696b} 236} & {\cellcolor[HTML]{598ac5} 9}\\ 

     \bottomrule
    \end{tabular}
    \end{center}
\end{table*}

%% file: tables/RQ2.tex
%!TEX root=bare_jrnl_compsoc.tex

\begin{table}[!h]
    \caption{Chi-squared test for New, Deleted, and Modified methods}
    \label{tab:rq2}
    \begin{center}
    %\begin{tabular}{c|l|r|r|r|r|l|l|r|r|r|r}
    \begin{tabular}{l|l|r|r|r}
    \toprule
    %\multicolumn{1}{c|}{Project} & \multicolumn{1}{c|}{} & \multicolumn{1}{c|}{New} & \multicolumn{1}{c|}{Deleted} & \multicolumn{1}{c|}{Modified} & \multicolumn{1}{c|}{Project} & \multicolumn{1}{c|}{} & \multicolumn{1}{c|}{New} & \multicolumn{1}{c|}{Deleted} & \multicolumn{1}{c}{Modified}\\
    
    \multicolumn{1}{l}{Project} & \multicolumn{1}{c|}{} & \multicolumn{1}{c|}{New} & \multicolumn{1}{c|}{Deleted} & \multicolumn{1}{c}{Modified}\\
    \hline

\multirow{2}{*}{\href{https://github.com/apache/accumulo}{Accumulo}}	&$\chi^2$&	\rnd{2}{61.553729}**	&	\rnd{2}{20.316330}**	&	\rnd{2}{174.749838}**	\\
&$\phi$&	\rnd{3}{0.368213}	&	\rnd{3}{0.324447}	&	\rnd{3}{0.595367}	\\	
\hline
\multirow{2}{*}{\href{https://github.com/apache/atlas}{Atlas}}	&$\chi^2$&	\rnd{2}{227.077192}**	&	\rnd{2}{75.858948}**	&	\rnd{2}{238.927466}**	\\
&$\phi$&	\rnd{3}{0.532108}	&	\rnd{3}{0.454643}	&	\rnd{3}{0.532061}	\\	
\hline
\multirow{2}{*}{\href{https://github.com/apache/beam}{Beam}}	&$\chi^2$&	\rnd{2}{252.505223}**	&	\rnd{2}{32.527232}**	&	\rnd{2}{371.667356}**	\\
&$\phi$&	\rnd{3}{0.495367}	&	\rnd{3}{0.227404}	&	\rnd{3}{0.571738}	\\	
\hline
\multirow{2}{*}{\href{https://github.com/apache/calcite}{Calcite}}	&$\chi^2$&	\rnd{2}{155.089902}**	&	\rnd{2}{2.488574}	&	\rnd{2}{301.032038}**	\\
&$\phi$&	\rnd{3}{0.403197}	&	\rnd{3}{0.086318}	&	\rnd{3}{0.555084}	\\	
\hline
\multirow{2}{*}{\href{https://github.com/apache/cayenne}{Cayenne}}	&$\chi^2$&	\rnd{2}{154.051256}**	&	\rnd{2}{29.401796}**	&	\rnd{2}{306.580336}**	\\
&$\phi$&	\rnd{3}{0.421282}	&	\rnd{3}{0.246725}	&	\rnd{3}{0.564821}	\\	
\hline
\multirow{2}{*}{\href{https://github.com/apache/cxf}{CXF}}	&$\chi^2$&	\rnd{2}{499.396553}**	&	\rnd{2}{18.521238}*	&	\rnd{2}{618.960321}**	\\
&$\phi$&	\rnd{3}{0.511604}	&	\rnd{3}{0.173964}	&	\rnd{3}{0.5552}	\\	
\hline
\multirow{2}{*}{\href{https://github.com/apache/deltaspike}{DeltaSpike}}	&$\chi^2$&	\rnd{2}{46.837465}**	&	\rnd{2}{9.184077}	&	\rnd{2}{68.504072}**	\\
&$\phi$&	\rnd{3}{0.483929}	&	\rnd{3}{0.336725}	&	\rnd{3}{0.568447}	\\	
\hline
\multirow{2}{*}{\href{https://github.com/apache/drill}{Drill}}	&$\chi^2$&	\rnd{2}{50.001846}**	&	\rnd{2}{11.657119}	&	\rnd{2}{405.693081}**	\\
&$\phi$&	\rnd{3}{0.240012}	&	\rnd{3}{0.172887}	&	\rnd{3}{0.67706}	\\	
\hline
\multirow{2}{*}{\href{https://github.com/apache/incubator-dubbo}{Dubbo}}	&$\chi^2$&	\rnd{2}{24.301076}**	&	\rnd{2}{9.214876}	&	\rnd{2}{118.007580}*	\\
&$\phi$&	\rnd{3}{0.314941}	&	\rnd{3}{0.258408}	&	\rnd{3}{0.675002}	\\	
\hline
\multirow{2}{*}{\href{https://github.com/apache/flink}{Flink}}	&$\chi^2$&	\rnd{2}{425.377553}**	&	\rnd{2}{19.242134}*	&	\rnd{2}{686.366318}**	\\
&$\phi$&	\rnd{3}{0.398549}	&	\rnd{3}{0.110567}	&	\rnd{3}{0.488267}	\\	
\hline
\multirow{2}{*}{\href{https://github.com/apache/flume}{Flume}}	&$\chi^2$&	\rnd{2}{95.412273}**	&	\rnd{2}{0.518831}	&	\rnd{2}{100.283919}**	\\
&$\phi$&	\rnd{3}{0.505087}	&	\rnd{3}{0.062458}	&	\rnd{3}{0.511034}	\\	
\hline
\multirow{2}{*}{\href{https://github.com/apache/giraph}{Giraph}}	&$\chi^2$&	\rnd{2}{74.008149}**	&	\rnd{2}{13.217305}	&	\rnd{2}{106.427315}**	\\
&$\phi$&	\rnd{3}{0.489396}	&	\rnd{3}{0.304021}	&	\rnd{3}{0.582186}	\\	
\hline
\multirow{2}{*}{\href{https://github.com/apache/jackrabbit}{Jackrabbit}}	&$\chi^2$&	\rnd{2}{289.095647}**	&	\rnd{2}{10.652703}	&	\rnd{2}{380.248458}**	\\
&$\phi$&	\rnd{3}{0.456709}	&	\rnd{3}{0.131504}	&	\rnd{3}{0.501983}	\\	
\hline
\multirow{2}{*}{\href{https://github.com/apache/jclouds}{jclouds}}	&$\chi^2$&	\rnd{2}{486.338150}**	&	\rnd{2}{30.429795}**	&	\rnd{2}{464.593322}**	\\
&$\phi$&	\rnd{3}{0.538842}	&	\rnd{3}{0.183572}	&	\rnd{3}{0.503586}	\\	
\hline
\multirow{2}{*}{\href{https://github.com/apache/knox}{Knox}}	&$\chi^2$&	\rnd{2}{154.247629}**	&	\rnd{2}{42.840849}**	&	\rnd{2}{165.542038}**	\\
&$\phi$&	\rnd{3}{0.559353}	&	\rnd{3}{0.474845}	&	\rnd{3}{0.559405}	\\	
\hline
\multirow{2}{*}{\href{https://github.com/apache/kylin}{Kylin}}	&$\chi^2$&	\rnd{2}{193.135450}**	&	\rnd{2}{45.058536}**	&	\rnd{2}{512.662558}**	\\
&$\phi$&	\rnd{3}{0.376983}	&	\rnd{3}{0.242852}	&	\rnd{3}{0.581906}	\\	
\hline
\multirow{2}{*}{\href{https://github.com/apache/metron}{Metron}}	&$\chi^2$&	\rnd{2}{22.992492}**	&	\rnd{2}{0.719311}	&	\rnd{2}{91.611991}**	\\
&$\phi$&	\rnd{3}{0.263959}	&	\rnd{3}{0.068567}	&	\rnd{3}{0.523725}	\\	
\hline
\multirow{2}{*}{\href{https://github.com/apache/myfaces}{MyFaces}}	&$\chi^2$&	\rnd{2}{198.435067}**	&	\rnd{2}{10.179724}	&	\rnd{2}{129.655143}**	\\
&$\phi$&	\rnd{3}{0.715144}	&	\rnd{3}{0.250675}	&	\rnd{3}{0.540995}	\\	
\hline
\multirow{2}{*}{\href{https://github.com/apache/nifi}{NiFi}}	&$\chi^2$&	\rnd{2}{360.675493}**	&	\rnd{2}{38.112054}**	&	\rnd{2}{189.469608}**	\\
&$\phi$&	\rnd{3}{0.695327}	&	\rnd{3}{0.386599}	&	\rnd{3}{0.495727}	\\	
\hline
\multirow{2}{*}{\href{https://github.com/apache/oozie}{oozie}}	&$\chi^2$&	\rnd{2}{34.553338}**	&	\rnd{2}{2.813912}	&	\rnd{2}{87.284096}**	\\
&$\phi$&	\rnd{3}{0.337138}	&	\rnd{3}{0.155749}	&	\rnd{3}{0.523085}	\\	
\hline
\multirow{2}{*}{\href{https://github.com/apache/openwebbeans}{OpenWebBeans}}	&$\chi^2$&	\rnd{2}{97.275810}**	&	\rnd{2}{21.013746}**	&	\rnd{2}{208.811955}**	\\
&$\phi$&	\rnd{3}{0.419791}	&	\rnd{3}{0.271062}	&	\rnd{3}{0.568101}	\\	
\hline
\multirow{2}{*}{\href{https://github.com/apache/pdfbox}{PDFBox}}	&$\chi^2$&	\rnd{2}{237.850190}**	&	\rnd{2}{27.779212}**	&	\rnd{2}{231.954512}**	\\
&$\phi$&	\rnd{3}{0.604916}	&	\rnd{3}{0.343816}	&	\rnd{3}{0.571574}	\\	
\hline
\multirow{2}{*}{\href{https://github.com/apache/pulsar}{Pulsar}}	&$\chi^2$&	\rnd{2}{288.134967}**	&	\rnd{2}{16.384578}*	&	\rnd{2}{212.824817}**	\\
&$\phi$&	\rnd{3}{0.622317}	&	\rnd{3}{0.254987}	&	\rnd{3}{0.519036}	\\	
\hline
\multirow{2}{*}{\href{https://github.com/apache/sis}{SIS}}	&$\chi^2$&	\rnd{2}{227.855978}**	&	\rnd{2}{35.032751}**	&	\rnd{2}{122.022382}**	\\
&$\phi$&	\rnd{3}{0.750072}	&	\rnd{3}{0.36021}	&	\rnd{3}{0.533324}	\\	
\hline
\multirow{2}{*}{\href{https://github.com/apache/storm}{Storm}}	&$\chi^2$&	\rnd{2}{133.336742}**	&	\rnd{2}{4.787651}	&	\rnd{2}{117.202648}**	\\
&$\phi$&	\rnd{3}{0.653729}	&	\rnd{3}{0.169827}	&	\rnd{3}{0.59868}	\\	
\hline
\multirow{2}{*}{\href{https://github.com/apache/tinkerpop}{TinkerPop}}	&$\chi^2$&	\rnd{2}{480.657984}**	&	\rnd{2}{30.516232}**	&	\rnd{2}{706.494142}**	\\
&$\phi$&	\rnd{3}{0.472604}	&	\rnd{3}{0.160882}	&	\rnd{3}{0.551243}	\\	
\hline
\multirow{2}{*}{\href{https://github.com/apache/zeppelin}{Zeppelin}}	&$\chi^2$&	\rnd{2}{100.304576}**	&	\rnd{2}{9.871353}	&	\rnd{2}{240.530245}**	\\
&$\phi$&	\rnd{3}{0.373505}	&	\rnd{3}{0.197139}	&	\rnd{3}{0.561465}	\\	

     \bottomrule
     \multicolumn{5}{l}{$^{\mathrm{**}}$ p $<$ 0.01, $^{\mathrm{*}}$ p $<$ 0.05}\\
    \end{tabular}
    \end{center}
 \end{table}

%% file: tables/RQ3.tex
%!TEX root=bare_jrnl_compsoc.tex

\begin{table}[!t]
    \caption{Adoption of Commit Guidelines and frequency of references to code quality issues in board meetings per project}
    \label{tab:rq3}
    \begin{center}
    \begin{tabular}{l|c|c}
        \toprule
        \multicolumn{1}{c|}{Project} & \multicolumn{1}{c|}{Commit Guidelines} & \multicolumn{1}{c}{Project Board Meetings}\\
        \midrule
        \href{https://github.com/apache/accumulo}{Accumulo} & NO & LOW\\
        \href{https://github.com/apache/atlas}{Atlas} & NO & LOW\\
        \href{https://github.com/apache/beam}{Beam} & NO & HIGH\\
        \href{https://github.com/apache/calcite}{Calcite} & YES & HIGH\\
        \href{https://github.com/apache/cayenne}{Cayenne} & NO & LOW\\
        \href{https://github.com/apache/cxf}{CXF} & YES & LOW\\
        \href{https://github.com/apache/deltaspike}{DeltaSpike} & YES & LOW\\
        \href{https://github.com/apache/drill}{Drill} & YES & HIGH\\
        \href{https://github.com/apache/incubator-dubbo}{Dubbo} & YES & LOW\\
        \href{https://github.com/apache/flink}{Flink} & YES & LOW\\
        \href{https://github.com/apache/flume}{Flume} & NO & LOW\\
        \href{https://github.com/apache/giraph}{Giraph} & NO & LOW\\
        \href{https://github.com/apache/jackrabbit}{Jackrabbit} & NO & HIGH\\
        \href{https://github.com/apache/jclouds}{jclouds} & YES & LOW\\
        \href{https://github.com/apache/knox}{Knox} & NO & LOW\\
        \href{https://github.com/apache/kylin}{Kylin} & YES & LOW\\
        \href{https://github.com/apache/metron}{Metron} & YES & HIGH\\
        \href{https://github.com/apache/myfaces}{MyFaces} & YES & LOW\\
        \href{https://github.com/apache/nifi}{NiFi} & YES & HIGH\\
        \href{https://github.com/apache/oozie}{oozie} & NO & LOW\\
        \href{https://github.com/apache/openwebbeans}{OpenWebBeans} & NO & HIGH\\
        \href{https://github.com/apache/pdfbox}{PDFBox} & YES & HIGH\\
        \href{https://github.com/apache/pulsar}{Pulsar} & YES & LOW\\
        \href{https://github.com/apache/sis}{SIS} & NO & HIGH\\
        \href{https://github.com/apache/storm}{Storm} & YES & LOW\\
        \href{https://github.com/apache/tinkerpop}{TinkerPop} & YES & LOW\\
        \href{https://github.com/apache/zeppelin}{Zeppelin} & YES & LOW\\
        \bottomrule
    \end{tabular}
    \end{center}
\end{table}

%% file: Discussion.tex
\section{Discussion}\label{sec:discussion}

In this section we first discuss in detail the findings presented in Section 4, attempting our own interpretation. Then, we list potential implications for researchers and practitioners.  

\subsection{Interpretations of the Results}

\subsubsection{Quality of new code vs. system quality (Descriptive Statistics)}

The descriptive statistics revealed that new code (in the form of new methods), exhibits quality (in terms of \TDd), that is higher compared to the quality of the systems in which the new methods are introduced. This finding was consistent in all examined projects. For many of the projects in the \asf ecosystem this could be linked to the observed declining trend of the project's \TDd, although this study cannot establish causality between cleaner new code and improvement of overall quality. It still remains to be studied how developers ensure that new code is cleaner in terms of \td. It might be a deliberate choice (e.g., in case the development team applies a Quality Gate to ensure zero or low number of violations for each new commit) or a general trend resulting from improvements in the employed processes and tools or simply the result of higher developer experience. 

It is also noteworthy that in a very large percentage of the analyzed commits, new methods had zero or very low \TDd. In particular, the findings from the selected projects, suggest that the overall system quality can improve (i.e. \TDd can decrease) over time if the \TDd of new code is systematically kept below the system's average. We caution, that this does not imply a quick fix to a system's quality (in terms of TD); if the existing code base is largely of low quality, the improvement that can be achieved through new commits is limited in the short term. Nevertheless, the improvement of {\TDd} in new code has merit in the sense that most systems evolve for years; thus, in the long run, systematically writing cleaner new code can yield substantial code improvement. This observation, emphasizes the importance of writing clean new code over the practice of refactoring. We also note that several empirical studies have shown that systematic, bad smells refactoring is uncommon in most \oss projects, rendering the practice of writing clean new code even more valuable. This result can be of particular worth, in the sense that writing clean code can be a best practice for limiting the \textit{software ageing} phenomenon \cite{parnas1994software} or the 7$^{th}$ of Lehman's laws of evolution which states that the quality of software deteriorates over time \cite{lehman1996laws}. 

\subsubsection{Relation among the contribution of new / deleted / modified methods and change in system's \TDd (RQ$_{1}$)}

The in-depth study on the association between the contribution of new / deleted / modified methods and the observed changes in the system's \TDd revealed some interesting patterns: (a) among all transitions where the system's \TDd has decreased and two or more types of changes were competing, the most frequent case was the addition of new methods that were better in terms of \TDd, implying that the introduction of clean new code coincided with an improvement in quality; (b) the contribution of method removal is rather mixed, which is reasonable, as the effect of method deletion on the system's \TDd depends on the quality of the removed code (the removed methods can be either high or low quality code); and (c) the contribution (positive or negative) of method modification coincides with the direction of change in the system's \TDd. For the latter case it should be emphasized that modifications refer to any type of changes to a method including adaptive and corrective maintenance.

The chi-square test of independence showed vividly that there is a statistically significant relationship between the direction of change in the system's \TDd and the contribution of new, deleted and modified code. Considering that code deletion is not usually performed on the basis of quality improvement but rather dictated by functionality-related reasons, the improvement of code quality is subsequently left up to the addition and modification of code. Method modification has a clear association with the overall \td: lowering the \TDd of a method during maintenance will reduce the system's \TDd and vice-versa. 
%However, throughout the evolution of the examined systems, we have observed both types of changes (modifications and additions) happening. 
On the other hand, cleaner new methods (which according to the descriptive statistics are very common) have a strong association to a decreasing system \TDd. These results, further emphasize the importance of writing high-quality new code and monitoring the introduced number of \TDIs in new code. To provide insight into how TD can be eliminated or avoided during software evolution, we provide two real examples: the first refers to code modification and the second to the introduction of clean new code.

The first example (\textbf{code modification}\textit{}) refers to a pull request in project \project{Dubbo} (\#3474) where the purpose of the change was to properly close resources after use, thereby eliminating an existing TD issue. The intention is also reflected on the title of the pull request: \lq Fix Not Properly Closed Resources\rq. The affected code, prior to the change, was:

\begin{lstlisting}[basicstyle=\small][language=Java]
try {
  UnsafeByteArrayInputStream is = new 
  UnsafeByteArrayInputStream((byte[]) args[i]);
 . . .
 } catch (Exception e) {
  . . .
 }
\end{lstlisting}

\noindent
and thus was missing a proper close call within a finally block, causing a Blocker issue according to \sq.
The change in the pull request targeted exactly that problem and fixed it by using the try-with-resources statement, that declares the resources to be closed after the program is finished (which is equivalent to using a finally block prior to Java SE 7). The corresponding code was modified to (note that the resource is declared within parentheses after the try keyword):

\begin{lstlisting}[basicstyle=\small][language=Java]
try(UnsafeByteArrayInputStream is = new 
  UnsafeByteArrayInputStream((byte[]) args[i])){
     . . .
   }
   catch (Exception e) {
     . . .
   }
\end{lstlisting}

\noindent
thereby, eliminating the abovementioned \sq issue.

The second example (\textbf{of clean new code introduction}\textit{}) refers to a pull request in the same project \project{Dubbo} when new methods are introduced in the new NettyClientHandler class. Up to that point, the code suffered from multiple TD issues related to improper handling of Exceptions, violating the major \lq Throwable and Error should not be caught\rq  rule 349 times. Noncompliant code examples are of the following form:

\begin{lstlisting}[basicstyle=\small][language=Java]
try {/* ... */} catch (Throwable t) {/* ... */}  
try {/* ... */} catch (Error e) {/* ... */} 
\end{lstlisting}

The rationale for this rule is that Throwable is the superclass of all errors and exceptions in Java, while Error is the superclass of all errors, which are not meant to be caught by applications. Catching either Throwable or Error will also catch OutOfMemoryError and InternalError, from which an application should not attempt to recover\footnote{\url{https://rules.sonarsource.com/}}.
Class NettyClientHandler in pull request \#630 is TD free and while it deals with exception handling in all of its methods, none of the methods violates the aforementioned rule.

\subsubsection{Code Quality Practices and Cleanness of New Code {(RQ$_{2}$)}}

We observed that projects in which Code Quality is often being discussed among the management team, exhibit a statistically significant higher percentage of commits with code that is cleaner than the existing codebase. These discussions are sometimes very explicit about the use of tools to measure code quality or the emphasis on cleaning up the code. As an example, in an Apache \project{PDFBox} board meeting of 2015, under \squote{Software Quality} it is mentioned that \emph{\squote{There is an ongoing effort to improve \project{PDFBox} based on the analysis of
different tools such as \sq, FindBugs and others}}. In a March 2019 meeting of project \project{Flink} under \squote{Status} it is noted that \emph{\squote{The release contains some new user-facing features plus a lot of internal cleanup and refactoring, fixing some long term issues ...}}.

Apache Project Management Committees (PMCs) are required to report on their project's health and status quarterly to the Board of Directors. We have observed in our study that projects with a certain level of size and complexity, code quality in general and maintainability in particular becomes a major concern. Guiding the hundreds of volunteers in open-source projects on how to commit high quality code can be facilitated by the use of tools/practices which essentially dictate a minimum threshold of quality that has to be reached before submitting code. We argue that such a \squote{clean new code} policy is also applicable to industrial projects as a means of sustaining, and even improving TD.

\subsection{Implications for Practitioners}\label{subsec:implicationsForResearchersAndPractitioners}

Regarding software developers, evidence from the presented case study suggests that new code can have a substantial impact on the quality of an evolving system. The fact that the contribution of new code has a strong association to the changes in the system's \TDd implies that writing clean new code can ensure, to a large extent, the gradual improvement of the overall system quality. Of course, code modifications that result in lower \TDd, either in the form of refactorings or as carefully applied maintenance, has also potential for improving code quality.

\input{tables/RepresentativeCommitGuidelines.tex}

In terms of software development strategies, we believe that using Quality Gates to enforce a predefined quality policy can be a simple, yet effective mechanism to manage \td in the long term. The findings on the second research question revealed that projects where code quality is a frequent topic in board meetings, are having higher chances to reduce their TD density through the improvement of new code.
The existence of explicit commit guidelines was not found to be significantly associated with the frequency of cleaner commits at the project level. However, it is noteworthy that several projects express explicit concerns about code quality in the commit guidelines offered to potential contributors. We list representative guidelines associated with code quality posted in the projects' websites in Table {\ref{tab:rq44}}. Such guidelines from well-known Apache projects can be considered as best practices and thus reused in other projects.

Ensuring that new code commits are as TD-free as possible, is a promising way of sustaining quality and avoiding quality degradation over time. Putting a Quality Gate in place is a relatively low-cost approach that sets an easy-to-manage, everyday goal for software developers, explicitly emphasizing code quality. 

% The analysis of Apache projects has shown that even the practice of posting 'how--to--commit' guidelines can be effective for improving the quality of new code

\subsection{Implications for Researchers}

With respect to researchers working in the area of \TDM the obtained evidence opens up further opportunities for studies on the impact of new code on quality. Research can focus on establishing whether systems with a degrading quality over time are associated with high {\TDd} of new code and vice-versa. Given that no project exhibits a monotonous trend in its quality over time, it would be reasonable to perform such studies after splitting the timeline of system history into periods with monotonous and statistically significant trends in their quality in terms of {\TDd}. Declaring that clean new code results in improving quality over time would emit an explicit message to software developers. 

Furthermore, the use of refactoring miners can be exploited to investigate which strategy (i.e., writing new code vs. applying regular refactorings) is more efficient for managing \td. Evidence on this central question would be highly relevant to software development teams considering both the effort that is associated with each type of quality-improvement approach as well as the attractiveness of each coding activity to developers. Moreover, it would be equally interesting to assess the quality of new code commits pertaining to specific change types, classified from the view point of the goal of change. Such classifications consider for example changes due to fault-fixing, feature addition, enhancements or general maintenance. \cite{hassan2009predicting, palomba2017scent}. Another research direction worth of investigation would be the study of \squote{ripple} effects of new code to the rest of the system. 

Further studies that compare projects that rely on tools such as \sq and others that do not, could reveal whether the use of such platforms leads to TD reduction. Moreover, it makes sense to investigate whether there is any relation between the experience of developers and the quality of new code that they introduce. Apart from the analysis of software artifacts we believe that it would be highly interesting to reach out to management boards of open source or industrial projects to obtain their own perception on \TD in new code and analyze their strategies for preventing its accumulation.

%% file: tables/RepresentativeCommitGuidelines.tex
%!TEX root=main.tex

\begin{table}[!t]
    \caption{Representative Commit Guidelines}
    \label{tab:rq44}
    \begin{center}
    \begin{tabular}{l|p{.75\linewidth}}
        \toprule
        \multicolumn{1}{c}{Project} & \multicolumn{1}{c}{Representative Commit Guidelines}\\
        \midrule
        % \href{https://github.com/apache/accumulo}{Accumulo} & \\
        % \href{https://github.com/apache/atlas}{Atlas} & \\
        % \href{https://github.com/apache/beam}{Beam} & \\
        \href{https://github.com/apache/calcite}{Calcite} & \textit{Trigger a Coverity scan ... and when it completes, make sure that there are no important issues.} \\
        % \href{https://github.com/apache/cayenne}{\textcolor{red}{Cayenne}} & \\
        \href{https://github.com/apache/cxf}{CXF} & \textit{Make use of both PMD and CheckStyle to enforce common coding conventions.} \\
        \href{https://github.com/apache/deltaspike}{DeltaSpike} & \textit{Follow project's formatting rules and always build and test your changes before you make pull requests} \\
        \href{https://github.com/apache/drill}{Drill} & \textit{Code should be formatted according to Sun's conventions, contributions should not introduce new Checkstyle violations, contributions should pass existing unit tests, and new unit tests should be provided to demonstrate bugs and fixes} \\
        \href{https://github.com/apache/incubator-dubbo}{Dubbo} & \textit{Dubbo uses code style that is almost in line with the standard java conventions and suggests the contributors to implement a few unit tests for a new feature or an important bugfix.} \\
        \href{https://github.com/apache/flink}{Flink} & \textit{Flink i) suggests the developers to comment as much as necessary to support code understanding, ii) provides guidelines for good design and software structure, iii) guidelines for good concurrency and threading and iv) suggestions for dependencies and modules} \\
        % \href{https://github.com/apache/flume}{Flume} & \\
        % \href{https://github.com/apache/giraph}{Giraph} & \\
        % \href{https://github.com/apache/jackrabbit}{Jackrabbit} & \\
        \href{https://github.com/apache/jclouds}{jclouds} & \textit{Contexts and APIs are thread-safe (or should! Otherwise it is an issue)} \\
        \href{https://github.com/apache/knox}{Knox} & \textit{Adding new service API support, the committer should give sufficient tests and documentation} \\
        \href{https://github.com/apache/kylin}{Kylin} & \textit{The changes MUST be covered by a unit test or the integration test, otherwise it is not maintainable} \\
        \href{https://github.com/apache/metron}{Metron} & \textit{Try-finally used as necessary to restore consistent state, Appropriate NullPointerException and IllegalArgumentException argument checks}  \\
        \href{https://github.com/apache/myfaces}{MyFaces} & \textit{Error and exception handling: If the exception is severe, but there is a chance to continue processing, a message with severity "error" or "warning" should be logged.}. \\
        \href{https://github.com/apache/nifi}{NiFi} & \textit{If an unexpected RuntimeException is thrown, it is likely a bug and allowing the framework to rollback the session will ensure no data loss} \\
        % \textit{Apache NIFI has a Review-Then-Commit (RTC) philosophy for handling all contributions.  Reviewers first ensure that the code applies and builds appropriately to the build, passes the code standards as established by the Maven profile "contrib-check."  From here, code is evaluated to ensure best practices within the NiFi framework are applied and, where applicable, that the user experience of interfacing with the contribution is consistent and any changes are backwards compatible.  This process may be iterative but works toward a final product that is then merged into the codebase. Furthermore, it uses checkstyle rules.} \\
        % \href{https://github.com/apache/oozie}{oozie} & \\
        % \href{https://github.com/apache/openwebbeans}{OpenWebBeans} & \\
        \href{https://github.com/apache/pdfbox}{PDFBox} & \textit{The new code should follow the project's coding conventions where possible.}\\
        \href{https://github.com/apache/pulsar}{Pulsar} & \textit{All code should have appropriate unit testing coverage. New code should have new tests in the same contribution. Bug fixes should include a regression test to prevent the issue from reoccurring.} \\
        % \href{https://github.com/apache/sis}{SIS} & \\
        \href{https://github.com/apache/storm}{Storm} & \textit{The most important is consistently writing a clear docstring for functions, explaining the return value and arguments. As of this writing, the Storm codebase would benefit from various style improvements..} \\
        \href{https://github.com/apache/tinkerpop}{TinkerPop} & \textit{A rich set of algorithms is an important goal for MLLib, scaling the project requires that maintainability, consistency, and code quality come first.} \\
        \href{https://github.com/apache/zeppelin}{Zeppelin} & \textit{The project follows Google's Java Code style and suggests the developers to use some formatting plugins to lint their code} \\
        \bottomrule
    \end{tabular}
    \end{center}
\end{table}

%% file: ThreatsToValidity.tex
%!TEX root=bare_jrnl_compsoc.tex

\section{Threats To Validity}\label{sec:threatstovalidity}

In this section, we present and discuss threats to the validity of the study, including threats to construct, external validity and reliability. The study does not aim at establishing the presence of cause-and-effect relationships, thus it is not concerned with internal validity.
%Some of these threats have been mitigated in the presented study design, while other threats can be addressed by further research on the contribution of new code to system quality. 

\subsection{Construct Validity}\label{subsec:constructvalidity}

Construct validity reflects how far the examined phenomenon is connected to the intended studied objectives. The main involved threat is related to the accuracy by which \td can be captured by static analysis tools such as \sq. Rule violations reported as \TDIs are obviously only one manifestation of actual code and design inefficiencies. The lack of any ground truth in \td measurement means that the accuracy of \sq, or any other \td tool, can hardly be validated. According to Martini et al. \cite{martini2018technical}, currently static analyzers (such as \sq) are used in industry to analyze the source code in search of \td. Only in few cases out of the respondents in their survey (15 companies) practitioners built their own metrics tools for checking (language-specific) rules or patterns that can warn the developers of the presence of \td. In a similar discussion, Yli-Huumo et al. \cite{yli2016software} found \sq to be the most used tool for TDM in the eight development teams that were involved in their case study. 

Furthermore, it is known that such tools are not capable of identifying architectural problems or other types of \td such as requirements, documentation, test or build debt. This threat however, is partially mitigated by the fact that \sq is one of the most widely used tools for \td identification and quantification. Moreover, we have used the same tool both for assessing the system's \TDd as well as the contribution of new / deleted / modified code on the \TDd change. Furthermore, the analysis of \TDd evolution is based mostly on a relative assessment of the changes rendering the measurement of absolute \td values less important for this type of study.
% The main threat related to construct validity stems from the fact that we do not investigate how the issues of the analyzed projects were eliminated (i.e. were they removed intentionally or it was a side effect).
% However, even in the presence of this threat the findings on deliberate or unintentional TD payback can be valuable since the focus is on the contribution of 
% Furthermore, we rely on \sq and its way in order to detect and estimation that effort that is required to pay back the \td. 

%However, even in the presence of this threat the findings on deliberate or unintentional TD payback can be valuable since the focus is on the relative frequency of each issue type rather than the absolute count of issue removals. An additional  threat stems from the fact that TD identification is performed on source code artifacts rather than the built version of each project. The lack of build information often leads to false positives. For example, a language specific issue (e.g. the omission of the diamond operator) might be reported for a revision in  which the rule does not apply (language version prior to Java 7).

%\subsection{Internal Validity}\label{subsec:internal}
In addition, we investigate the effect of code insertion, deletion and modification on changes in the system's \TDd. Generally, these three types correspond to all possible changes in terms of code. However, we consider only \TD that can be mapped to methods, thus ignoring TD which might occur at the level of classes or files. \sq reports violations at the class/file level; however, tracking the types of changes that can introduce or remove such violations requires a different study set-up. More importantly, we consider the co-occurrence of changes (e.g. an increase of \TDd by new code and an increase in the system's \TDd). Thus, we do not capture the underlying causes of variation in the number of \TD issues (such as the introduction of a new library or framework, the implementation of new functionality, etc.). 

\subsection{Reliability}\label{subsec:reliability}
Reliability reflects whether the study has been conducted and reported in a way so that others can replicate it and reach the same results. To mitigate this threat, the study protocol is extensively described in Section \ref{sec:caseStudyDesign} explicitly listing all data collection and analysis steps. It should be emphasized that data has been subject only to automated analysis with no subjective interpretation from the researchers; therefore, researcher bias has been avoided. A replication package is available with all available data to allow for an independent replication of the investigation. Replication to other software ecosystems like the ones by Google, Android and Salesforce is deemed particularly important for validating the findings on the effectiveness of clean new code.

% Finally, there is a \textcolor{red}{replication kit} with all the data and tool available, so other researchers can download, use, validate, and even enhance them.

\subsection{External Validity}\label{subsec:externalvalidity}

External validity examines the applicability of the findings in other settings, e.g. other software projects, other programming languages and possibly other \td identification tools. We have focused only on Java \oss projects that use maven as a build tool. This limits the ability to generalize the findings to other projects. The fact that the study focuses on \TSEsyscount projects of the \asf which are highly active and popular among software developers partially mitigates threats to generalization. Nevertheless, the focus on a specific software ecosystem is still a source of threat to the generalizability of the results. Last but not least, the introduction of TD through new code has been analyzed in this study only on open source projects. In industry, projects are characterized by a very tight schedule; this means that our findings cannot be generalized to industrial systems. Thus, replication studies on the effect of new code on the evolution of \td are needed to strengthen the validity of the derived conclusions to industrial systems or systems that use different programming languages and environments. 

%% file: Conclusion.tex
%!TEX root=bare_jrnl_compsoc.tex

\section{Conclusions}\label{sec:conclusion}

The Technical Debt metaphor is usually associated with degrading quality trends in software systems. However, not all systems \textit{age} over time. In this paper we have made an attempt to shed light into the drivers of \td change, by focusing on the contribution of new, deleted and modified code. In particular, we have performed an empirical study on the entire history of \TSEsyscount \os projects by analyzing the changes, at method level, for each individual commit. By mapping \TDIs to new, deleted, and modified methods we have been able to investigate the association between the types of changes and the variation in the system's \TDd.

The results revealed that the quality of new code in terms of its \TDd, is, for the majority of the revisions, higher than the quality of the system in which the code is introduced. Moreover, among new, deleted, and modified code, the contribution of code modification exhibits a strong association to the change in the system's \TDd, followed by the contribution of new code. The contribution of new code is more profound for the transitions in which the quality of the system improved. More specifically, it was found that adding new code that is \textit{cleaner} than the existing codebase coincides very frequently with a reduction in \TDd of the system. The same association, has also been observed for the contribution of method modification to the change of the total \TDd. Finally, we have found indications that projects in which code quality is often discussed in their board meetings, exhibit a higher frequency of cleaner code commits. 

The findings of this study suggest that writing code that has fewer \TDIs than the host code, can prove a very efficient and low-cost approach for managing TD. Applying Quality Gates to ensure that each commit yields fewer violations than the average, essentially leads to an improving quality trend, thereby reversing the software ageing phenomenon. Further studies can reveal whether writing clean new code offers a better cost/benefit ratio than the widely studied strategy of software refactoring.